\documentclass[11pt,oneside,letterpaper]{article}
\usepackage{amssymb}
\usepackage{amsmath}
\usepackage{amscd}
\usepackage{graphicx}
\usepackage{setspace}
\usepackage{fancyhdr}
\usepackage{marvosym}
\usepackage{ifpdf}
\usepackage{graphicx}
\usepackage{comment}

 \def\p{\partial}

\def\beq{\begin{equation}}
\def\eeq{\end{equation}}
\def\y{\alpha }

\newcommand{\bea}{\begin{eqnarray}}
\newcommand{\eea}{\end{eqnarray}}
\newcommand{\be}{\begin{equation}}
\newcommand{\ee}{\end{equation}}

\newcommand{\bi}{\begin{itemize}}
\newcommand{\ei}{\end{itemize}}

\newcommand{\qed}{\nobreak \ifvmode \relax \else
      \ifdim\lastskip<1.5em \hskip-\lastskip
      \hskip1.5em plus0em minus0.5em \fi \nobreak
      \vrule height0.75em width0.5em depth0.25em\fi}

\numberwithin{equation}{section}
\allowdisplaybreaks  

\begin{document}

\vspace*{1.5cm}
\begin{center}
{  \Large{\textsc{Incompressible Fluids of the de Sitter Horizon \newline\newline and Beyond}} }
\vspace*{1.5cm}

Dionysios Anninos$^{\heartsuit}$, Tarek Anous$^{\clubsuit}$,  Irene Bredberg$^{\clubsuit}$ and Gim Seng Ng$^{\clubsuit}$
\vspace*{1cm}

{\it $^\heartsuit$ Department of Physics, Stanford University, \\ Stanford, CA 94305-4060, USA}
\vspace*{0.5cm}

{\it $^\clubsuit$ Center for the Fundamental Laws of Nature, \\ Harvard University, Cambridge, MA 02138, USA}
\end{center}
\vspace*{1.5cm}

\begin{abstract}

There are (at least) two surfaces of particular interest in eternal de Sitter space. One is the timelike hypersurface constituting the lab wall of a static patch observer and the other is the future boundary of global de Sitter space.
We study both linear and non-linear deformations of four-dimensional de Sitter space which obey the Einstein equation.
Our deformations leave the induced conformal metric and trace of the extrinsic curvature unchanged for a fixed hypersurface. This hypersurface is either timelike within the static patch or spacelike in the future diamond. We require the deformations to be regular at the future horizon of the static patch observer. For linearized perturbations in the future diamond, this corresponds to imposing incoming flux solely from the future horizon of a single static patch observer.
When the slices are arbitrarily close to the cosmological horizon, the finite deformations are characterized by solutions to the incompressible Navier-Stokes equation for both spacelike and timelike hypersurfaces. We then study, at the level of linearized gravity, the change in the discrete dispersion relation as we push the timelike hypersurface toward the worldline of the static patch. Finally, we study the spectrum of linearized solutions as the spacelike slices are pushed to future infinity and relate our calculations to analogous ones in the context of massless topological black holes in AdS$_4$.

\end{abstract}

\newpage

\tableofcontents


\section{Introduction -- $\mathcal{I}^+$/Static Patch Schizophrenia}

The observation that the universe is dominated in energy by a  positive cosmological constant \cite{Perlmutter:1998np,Riess:1998cb} suggests that it may evolve into an asymptotically de Sitter era in the future. In such a context, any surviving observers will find themselves surrounded by a cosmological horizon whose size is set by the value of the cosmological constant. The nature of this horizon is rather enigmatic given that an observer can never reach her cosmological horizon -- it is always at a finite fixed distance away from her. Thus, unlike a black hole, the static patch horizon is not localized in some finite region of space. On the other hand, it behaves classically, and to an extent quantum mechanically, very much like a black hole horizon. For instance there is a temperature and entropy associated with the cosmological horizon \cite{Gibbons:1977mu}.

One of the main challenges confronted by theorists is to uncover the nature of the holographic principle in the context of asymptotically de Sitter universes. One may be inclined to propose that de Sitter holography should only describe a single patch, given that this is the region of space accessible to a single observer \cite{Goheer:2002vf,Banks:2003cg,Parikh:2004wh,Banks:2006rx,Banks:2002wr,Susskind:2011ap,Parikh:2008iu,Castro:2011xb,Alishahiha:2004md,Silverstein:2003jp,Anninos:2011jp,Anninos:2011af}. In fact, the observer's worldline in the static patch resembles in many ways the boundary of anti-de Sitter space (in the presence of an eternal black hole) and may constitute the ultraviolet regime of the fundamental description \cite{Alishahiha:2004md,Anninos:2011jp,Anninos:2011af}, perhaps in a way related to matrix theory. In such a case, the dynamics of the cosmological horizon would constitute the deep infrared behavior of the putative worldline theory very much like the near horizon dynamics of a black hole in anti-de Sitter space constitutes the deep infrared behavior of the dual conformal field theory. Such behavior rather universally  takes the form of fluid dynamics.

There have been several attempts to relate general relativity to fluid mechanics dating back to the 1970s with the black hole membrane paradigm \cite{thesDamour,oldDamour,Price:1986yy} (see \cite{Khoury:2006hg} for an application to de Sitter space). The membrane paradigm focuses on the observation that the equations governing the dynamics of horizon surfaces in general relativity can be written in a form analogous to that of the Navier-Stokes equation of fluid mechanics. However, whilst finding a striking analogy, the central equation of the membrane paradigm is often referred to as the Damour-Navier-Stokes equation, highlighting the fact that it differs from the Navier-Stokes equation in key ways. Building on this, recent papers \cite{Bredberg:2010ky,B11b,Bredberg:2011xw} constructed a setup where near horizon dynamics in gravity precisely relates the Einstein equation to the incompressible Navier-Stokes equation. These studies were also inspired by analyses of connections between gravity and fluid mechanics in the context of the AdS/CFT correspondence \cite{Policastro:2001yc,minw,Bhattacharyya:2008jc} and the low energy limit of the dual field theory. Given the striking similarities between the thermodynamics of a black hole horizon and a cosmological horizon, it is natural to extend such a fluid/gravity correspondence to include spacetimes with a cosmological horizon.

Yet another natural location for a definition of quantum gravity in de Sitter space is the spacelike boundary at future infinity, known as $\mathcal{I}^+$ \cite{Hartle:1983ai,Strominger:2001pn,Hull:1998vg,Witten:2001kn,Maldacena:2002vr,Harlow:2011ke,Anninos:2010zf}. The `metaobservables' on $\mathcal{I}^+$ are given by correlators between causally disconnected points. In a sense, we are metaobservers of the `would be' $\mathcal{I}^+$ of the inflationary de Sitter era. Such metaobservables are conjectured to constitute the correlation functions of a lower dimensional non-unitary Euclidean conformal field theory. Recently, an exact example of this dS/CFT correspondence was proposed in \cite{Anninos:2011ui}. It is not at all clear how the physics of the static patch observer is captured by the theory at $\mathcal{I}^+$. The static patch observer can at most observe a single point (or tiny region) of $\mathcal{I}^+$ where her worldline intersects the future boundary. Thus, from the static patch observer's point of view, fixing the geometry outside her future horizon is akin to fixing a gauge since it will never affect the physics she observes \cite{Anninos:2011jp}. In particular, one might envision fixing the geometry near $\mathcal{I}^+$ by fixing the data on a spacelike slice in the future diamond and allowing only flux originating from a single static patch to come through. This is somewhat analogous to the boundary condition that there is no incoming flux from the past horizon of a black hole. It may then be speculated that the existence of a finite number of `holographic reconstructions' of a static patch observer in the dual CFT at $\mathcal{I}^+$ would be a manifestation of the finiteness of the de Sitter entropy.\footnote{This may ultimately be related to the mysterious use of the Cardy formula for `counting' the de Sitter entropy in some lower dimensional examples \cite{Bousso:2001mw,Anninos:2011vd}.}


After reviewing the classical geometry of de Sitter space,
the first part of this paper will explore some of the classical features of the cosmological horizon as viewed by an observer in a purely de Sitter universe -- the static patch observer. We examine the Einstein equation both linearly and non-linearly and uncover that the solutions are characterized by solutions to the incompressible Navier-Stokes equation on a two-sphere.\footnote{As in \cite{Bredberg:2011xw}, we analyze the metric through the first three orders in a near-horizon expansion. A generalization of the all-orders proof of \cite{Compere:2009} might be possible in our case, but we will not attempt to do so herein.} This same equation recently appeared in the context of the Schwarzschild black hole \cite{Bredberg:2011xw} and requires the velocity field $v_i(\tau,\Omega^j)$ where $\Omega^i=\{\theta,\phi\}$ and the pressure $P(\tau,\Omega^j)$ to satisfy
\begin{equation}\label{navierstokesfluid}
\partial_\tau v^i + \nabla_{S^2}^i P + v_j \nabla^j_{S^2} v^i - \nu \left( \nabla^2_{S^2} v^i + R^{i}_{j} v^j \right) = 0~, \quad \nabla_{S^2}^i v_i = 0~
\end{equation}
where $\nu$ is the viscosity. Indices are raised and lowered with respect to the round metric $g_{ij}$ on the $S^2$ of radius $r_S$ for which $R_{ij} (= g_{ij}/r_S^2)$ is the Ricci tensor. At the linearized level, this is done by imposing Dirichlet boundary conditions on a timelike surface arbitrarily close to the cosmological horizon and the absence of incoming flux from the past horizon of the static patch. These boundary conditions resemble the solipsistic boundary conditions of \cite{Anninos:2011af}, which allow for an examination of the isolated static patch dynamics, unperturbed by external sources from the past horizon. We find that the linearized solutions must obey the dispersion relation of the incompressible, linearized (pressureless) Navier-Stokes equation~(\ref{linnavierstokes}).
At the non-linear level, again in a near cosmological horizon expansion, we impose (conformal) Dirichlet boundary conditions on a timelike slice and regularity of the solutions as they approach the future horizon. By (conformal) Dirichlet boundary conditions, we mean analysing perturbations which leave the induced geometry on a fixed timelike hypersurface of constant extrinsic curvature unchanged up to a conformal factor.\footnote{Henceforth, in the non-linear analysis, we will refer to these boundary conditions as Dirichlet boundary conditions.}
Then, we comment briefly on the possibilities of deforming this non-linear fluid by placing a small black hole at the origin of the static patch. In an attempt to connect our fluid dynamical modes to the analogous excitations of the worldline, which are the quasinormal modes, we return to the linearized analysis to study how the linearized dispersion relation varies as we push the surface from the cosmological horizon to the worldline.

In the second part of the paper we make some mathematical observations about spacelike slices foliating the region outside the future horizon of the static patch. We examine the behavior of linearized solutions to the Einstein equation near, but outside, the future cosmological horizon. Our solutions are subjected to Dirichlet boundary conditions on a fixed spacelike surface and to contain incoming flux solely from a single static patch observer. We find a discrete set of modes obeying the dispersion relation of the linearized Navier-Stokes equation, where the time coordinate has become the non-compact spacelike coordinate moving us along the spacelike slice. The non-linear solutions to the Einstein equation which satisfy Dirichlet boundary conditions on the spacelike slice and which are regular at the horizon from which flux is coming, are indeed characterized by solutions to the incompressible Navier-Stokes equation. The Navier-Stokes equation uncovered here on the spacelike slice is equivalent to that discussed in the context of the timelike surface, except that the sign of the viscosity is flipped. We end by noting that the setup of the problem in this future diamond of de Sitter space, and in particular the pole structure at $\mathcal{I}^+$, is connected by an analytic continuation to analogous problems in Lorentzian AdS$_4$ with hyperbolic slicing.


\section{Geometry and Framework}

In what follows we will study the geometries of several patches of de Sitter space pertinent to our analysis. Instead of the global patch of de Sitter space containing the past and future infinities, denoted by $\mathcal{I}^-$ and $\mathcal{I}^+$, we will focus on patches that are more suited to the description of local observers.

\subsection{The Static Patch}

The four-dimensional static patch metric solves the Einstein equation in the presence of a cosmological constant $\Lambda > 0$,
\begin{equation}
{\cal G}_{\mu\nu} \equiv G_{\mu\nu} + \Lambda g_{\mu\nu}=0
\end{equation}
 and is given by:
\begin{equation}\label{static}
ds^2 = - \left( 1 - (r/\ell)^2 \right) dt^2 +  \left( 1 - (r/\ell)^2 \right)^{-1} dr^2 + r^2 d\Omega^2_2~,
\end{equation}
where $r \in [0,\ell]$, $t \in \mathbb{R}$ and $d\Omega_2^2$ is the round metric on $S^2$.  The quantity $\ell$ is the de Sitter length and is related to the cosmological constant as $\Lambda = + 3/\ell^2$. The above metric covers a quarter of the global de Sitter geometry, it describes the intersection of the future and past causal diamonds of a constant $r$ worldline beginning at $\mathcal{I^-}$ and ending at $\mathcal{I}^+$. We call this the Southern patch of de Sitter space.

One notices that $r = \ell$ corresponds to a cosmological event horizon, beyond which events are forever out of causal contact from the Southern observer. The Killing vector $\partial_t$ becomes null at $r = \ell$ and the above coordinate system breaks down.

The Southern patch can be smoothly connected to another region covering an additional quarter of de Sitter space, by continuing the above metric to $r \in [\ell,\infty]$. For $r > \ell$, $t$ becomes a spacelike coordinate and $r$ becomes timelike. We can consider gluing two such regions, one behind the past cosmological horizon, known as the past diamond containing $\mathcal{I}^-$, and the other beyond the future cosmological horizon, known as the future diamond containing $\mathcal{I}^+$.

\begin{figure}[ht!]
\centering
\includegraphics[width=0.4\textwidth]{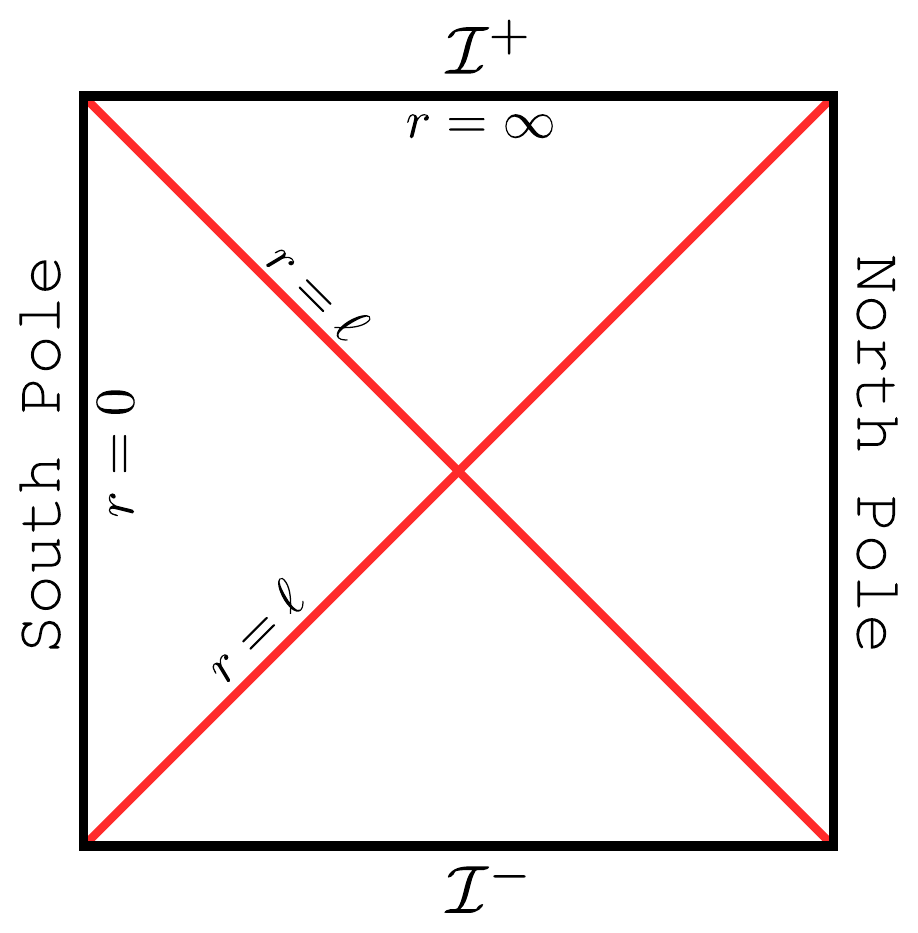}
\caption{Penrose diagram of de Sitter space indicating the various static patches and future/past diamonds.}\label{dSpenrose}
\end{figure}

The remaining quarter of the global de Sitter space is given by an additional static patch system known as the Northern patch. The Southern and Northern patches each intersects $\mathcal{I}^\pm$ at a single point. In figure~\ref{dSpenrose} we demonstrate the several patches discussed above in a Penrose diagram.

\subsection{Null Foliations}

It will be convenient to introduce an additional coordinate system which smoothly covers both the Southern patch and the future diamond. This is achieved by the following coordinate transformation:
\begin{equation}
\ell du = dt - \frac{dr}{ \left( 1 - (r/\ell)^2 \right)}~, \quad\quad v = \frac{r}{\ell}~,
\end{equation}
leading to the metric
\begin{equation}\label{null}
\frac{ds^2}{\ell^2} = -(1-v^2)du^2 - 2du dv + v^2 d\Omega^2_2~.
\end{equation}
Up to a constant time shift we find $u\ell = t -\ell \tanh^{-1}r/\ell$. Constant $u$ surfaces are null lines emanating from the origin at $v = 0$ and ending at $\mathcal{I}^+$ where $v = \infty$. The norm of the Killing vector $\partial_u$ changes sign at $v = 1$.
\begin{figure}[ht!]
\centering
\includegraphics[width=0.4\textwidth]{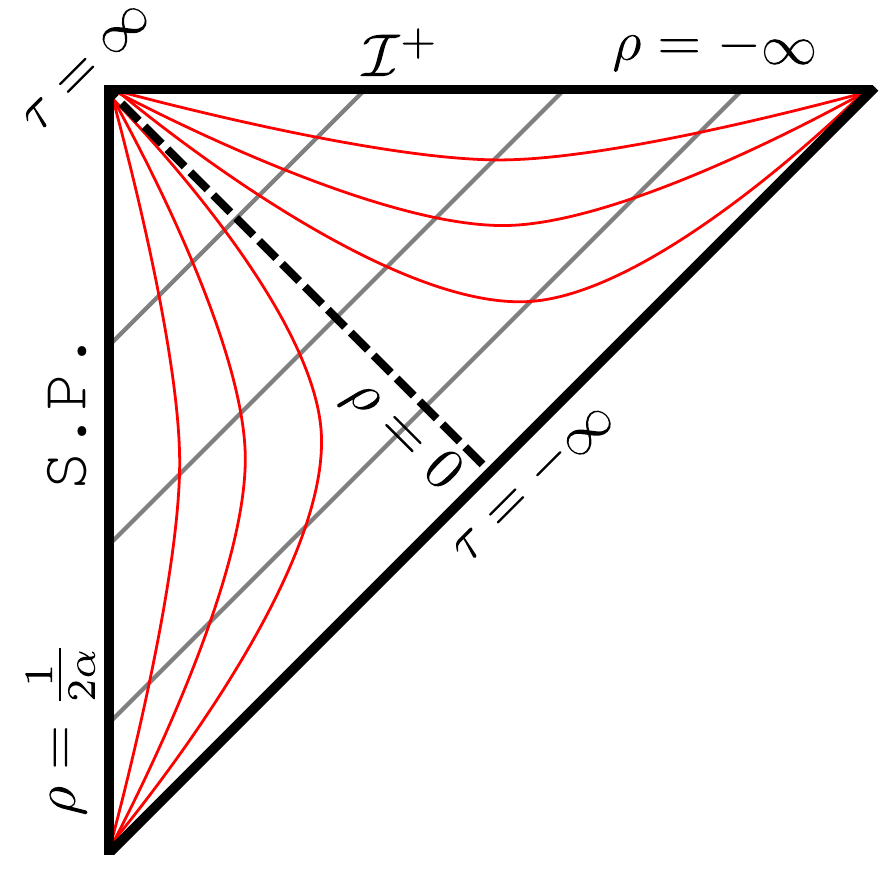}
\caption{Penrose diagram of de Sitter space indicating constant $\rho$ (red) and $\tau$ (gray diagonal) slices.}\label{rhotau}
\end{figure}

\subsection{Approaching the Horizon}

Finally, we would like to introduce a dimensionless parameter $\alpha > 0$ allowing us to approach the cosmological horizon. In order to achieve this, we rescale time to $u = \tau/2\alpha$ and define $\rho = (1 - v)/2\alpha$. As we take the limit $\alpha \to 0$, we redshift time and for any finite $\rho$, $v$ will be forced to lie near the cosmological horizon. The metric is given by: 
\begin{equation}\label{nearcosf}
\frac{ds^2}{\ell^2} =  \left( -\frac{\rho}{\alpha} +\rho^2  \right)d\tau^2 + 2d\tau d\rho + \left(  1 - 2\alpha\rho \right)^2 d\Omega_2^2~.
\end{equation}
The coordinate range of $\rho$ covering the Southern patch is given by $\rho \in [0,1/2\alpha]$ and the norm of $\partial_u$ vanishes at $\rho = 0$. The constant $\rho$ and $\tau$ surfaces are shown in figure~\ref{rhotau}.

As opposed to the Schwarzschild case, where a similar expansion would continue for indefinite powers of $\alpha$, the above expansion terminates at order $\mathcal{O}(\alpha^3)$. This is due to the absence of a term $\sim 2M/r$ in the $g_{\tau\tau}$ component. We could of course add such a term, which would correspond to introducing a small mass or black hole centered at the origin of the static patch.

\section{Incompressible Fluids}

Having specified the geometry relevant to our problem, we proceed to discuss the nature of perturbations solving the Einstein equation with positive $\Lambda$ near the cosmological horizon. We begin with a linearized analysis.

\subsection{Linearized Analysis}

Linearized gravity about spherically symmetric spaces with non-zero cosmological constant was examined in \cite{Kodama:2000fa,Kodama:2003jz}. The two gravitational degrees of freedom transform as a (divergenceless) vector and a scalar under the $SO(3)$ symmetry of the $S^2$. There is no transverse-traceless tensorial spherical harmonic for a two-sphere. Let us work in a gauge where $\delta g_{ij}=0$ for $x^i\in\{ \Omega \}$. The metric vector perturbations can be expressed as:
\begin{eqnarray}\label{metriccomps}
\delta g_{it} &=&  \mathcal{V}_i \times \left(  1 - (r/\ell)^2 \right) \left(  1 + r\partial_r  \right)\Phi_v~,\\
\delta g_{ir} &=&  \mathcal{V}_i \times \frac{r}{\left( 1-(r/\ell)^2 \right)} \partial_t \Phi_v~.
\end{eqnarray}
The vector spherical harmonic $\mathcal{V}_i$ satisfies the following relations on the unit two-sphere:
\begin{equation}\label{vecharmoniceq}
\left( \nabla^2_{S^2} + k_V^2 \right) \mathcal{V}_i = 0~, \quad \nabla^i_{S^2} \mathcal{V}_i = 0~,
\end{equation}
with eigenvalues are $k_V^2 = l(l+1)-1$ and $l = 1,2,\ldots$ The master field $\Phi_v$ obeys the master equation:
\begin{equation}
\left( \nabla^2_{g^{(2)}} - \frac{ l (l+1)}{r^2}  \right) \Phi_v = 0~,
\end{equation}
where $g^{(2)}$ corresponds to the two-dimensional de Sitter static patch. A similar result holds for the scalar perturbations, which we discuss in appendix \ref{scalar}.

The solutions to the above equation were analyzed in \cite{Anninos:2011jp} and are found to be hypergeometric functions. For our purposes we would like to obtain the linearized solutions in the null coordinate system (\ref{null}). Assuming a Fourier decomposition in time, $\Phi_v = e^{-2 i \alpha \omega  t(\tau,\rho)/\ell}\varphi_v(\rho)$, the equation of motion becomes:
\begin{multline}
\left( 4\rho^2\left( 1 - \alpha \rho \right)^2 \partial_\rho^2 + 4\rho \left( 1 - \alpha \rho \right)\left(  1 - 2\alpha\rho \right) \partial_\rho + \right. \\  \left.  \frac{4\alpha^2 ( 1 - 2\alpha\rho )^2 \omega^2 - 4\alpha \rho \left( 1 - \alpha \rho \right) \left( k_V^2 + 1  \right) }{(1-2\alpha\rho)^2} \right) \varphi_v = 0~.
\end{multline}
The two linearly independent solutions for $l>1$ are given by:
\begin{eqnarray}\label{waves}
\varphi_v^{out}&=& \rho^{-i\alpha\omega} \; {_2}F_1 \left[ a_1,b_1; c_1; \alpha \rho (-1+2\alpha\rho)^{-1} \right]  \;  \left( 1 - \alpha \rho \right)^{-i\alpha\omega} \left(  1 - 2\alpha \rho \right)^{2i\alpha\omega} ~,\\\label{waves2}
\varphi_v^{in} &=&  \rho^{+i \alpha\omega}\; {_2}F_1 \left[  a_2,b_2;c_2; \alpha \rho (-1+2\alpha\rho)^{-1} \right] \; (1-\alpha\rho)^{-i\alpha\omega}~,
\end{eqnarray}
with:
\begin{eqnarray}
a_1 &=& -l-2i\alpha\omega~, \quad b_1 = 1+l-2i\alpha\omega~, \quad c_1 = 1 - 2i\alpha\omega~;\\
a_2 &=& - l~, \quad b_2 = 1+l~, \quad c_2 = 1 + 2i\alpha\omega~.
\end{eqnarray}
The superscripts `out' and `in' indicate that the mode is purely outgoing at the future horizon or purely incoming from the past horizon. The above expressions are linearly independent solutions for $(c_i - a_i - b_i)  = 2i\alpha\omega \notin \mathbb{Z}$ (see \cite{stegunch15}). In the case where  $(c_i - a_i - b_i)  = 2i\alpha\omega$ is an integer, logarithmic solutions will appear. Given that $a_2$ and $b_2$ are integers $\varphi_v^{in}$ is in fact a finite polynomial for $2i\alpha\omega \notin \mathbb{Z}$, as it can be shown that the hypergeometric series terminates. For $l = 1$, the linearized perturbations become time independent and are like the introduction of a small amount of angular momentum (we discuss this case in appendix \ref{smallrot}).

The linearized purely outgoing metric components (\ref{metriccomps}) in the $(\tau,\rho)$-coordinate system become:
%
\begin{align}\label{metricII}
\delta g^{out}_{i \tau} &= 2\mathcal{V}_i \times e^{-i\omega\tau} \; \rho^{i\alpha\omega+1}(1 - \alpha\rho)^{-i\alpha\omega+1} \left( 1- \frac{(1-2\alpha\rho)}{2\alpha} \partial_\rho \right)\varphi_v^{out}~,\\
\delta g^{out}_{i \rho} &=-2\alpha\mathcal{V}_i \times e^{-i\omega\tau} \;  \left( \frac{1-\alpha\rho}{\rho}\right)^{-i\alpha\omega}\left[ \left(1-\frac{i\omega(1-2\alpha\rho)}{2\rho(1-\alpha\rho)}\right)- \frac{(1-2\alpha\rho)}{2\alpha} \partial_\rho \right]\varphi_v^{out}~.
\end{align}
Both $\delta g^{out}_{i \tau} $ and $\delta g^{out}_{i \rho} $ are regular at the future horizon $\rho = 0$.


\subsection{Linearized Fluid Modes}

Having written down the linearized solutions, we now discuss the choice of boundary conditions. We impose Dirichlet boundary conditions on a given timelike hypersurface at some fixed $\rho$, and without loss of generality, we choose the location of this timelike hypersurface to be at $\rho = 1$. Taking $\alpha \to 0$ pushes this hypersurface arbitrarily near the cosmological horizon and thus allows us to probe the near horizon dynamics.

\begin{figure}[ht!]
\centering
\includegraphics[width=0.2\textwidth]{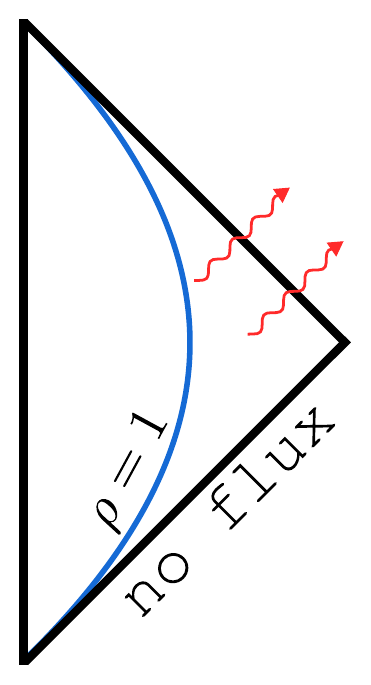}
\caption{Our boundary conditions for the linearized modes are such that the induced metric on the $\rho=1$ slice is unchanged and there is no incoming flux from the past horizon.}\label{staticpatchbc}
\end{figure}

Our particular Dirichlet boundary condition, shown in figure~\ref{staticpatchbc}, is that the linearized perturbations are purely outgoing and leave the intrinsic geometry of the $\rho = 1$ hypersurface unchanged.\footnote{This is the simplest choice of Dirichlet boundary conditions and thus allows for a clear analysis. In general, we can choose more involved Dirichlet boundary conditions on the $\rho = 1$ hypersurface.}
Imposing $\delta g_{i\tau}^{out} (\rho = 1) =  0$ enforces a discrete dispersion relation, which to leading order in $\alpha$ is given by:
\begin{equation}\label{fluid}
\omega_{f}= - i \left( l(l+1) - 2\right)~, \quad l = 1,2,\ldots
\end{equation}
We interpret these linearized modes as fluid modes of the velocity field $v^i$ of the incompressible, linearized (pressureless) Navier-Stokes equation on a sphere:
\begin{equation}\label{linnavierstokes}
\partial_\tau v_i  = \nu \left(  \nabla_{S^2}^2 v_i + R_{ij}v^j \right)~, \quad \nabla_{S^2}^i v_i = 0~,
\end{equation}
where the viscosity $\nu = 1$. 
The incompressibility of the fluid is equivalent to the vanishing divergence of $\mathcal{V}_i$,
which can be seen by identifying $v_i \sim e^{-i\omega \tau}\mathcal{V}_i$.
  We further note that the explicit modes (\ref{metricII}) with $\omega = \omega_f$ decay in time and are regular at the future horizon $\rho = 0$.\footnote{If these modes are taken back in time to $t \to -\infty$ they diverge and the perturbative solution is no longer reliable. As usual we only consider wavepackets of the linearized solutions which are finite for all asymptotia.}

By arguments similar to those in \cite{Bredberg:2010ky}, one expects that the result $\nu = 1$ corresponds to a ratio of shear viscosity to entropy density which is $1/4\pi$. This suggests that the incompressible fluid we have found near the de Sitter horizon shares this feature with the fluids found near the Schwarzschild, Rindler and planar AdS black hole horizons \cite{Bredberg:2010ky,B11b,Bredberg:2011xw,Kovtun:2004de} (see also \cite{Khoury:2006hg}).

There is also a decoupled set of scalar excitations which transform as scalars under the $SO(3)$ of the $S^2$.
However, the incompressibility condition implies that we could only consider the spherically symmetric scalar mode, which reduces the fluid vector field to a trivial one(see appendix \ref{scalar}).
We will not consider such modes in what follows and simply set them to zero in the linearized analysis.

\subsection{Non-linear Analysis}

Having analyzed the linearized case, we now turn to the question of non-linear deformations. The analysis follows directly the Schwarzschild case analyzed in \cite{Bredberg:2011xw}.\footnote{It should be noted that we have presented a more complete linearized analysis than would be possible for the Schwarzschild case, given the existence of exact linearized solutions in $dS_4$.} In particular, in this non-linear analysis, we impose the (conformal) Dirichlet boundary conditions on the hypersurface as in \cite{Bredberg:2011xw}.

To be precise, we will consider the following finite deformation of the static patch geometry as an expansion in $\alpha$:\footnote{When writing out the metric (\ref{timenonlin}) we have omitted the metric components $g_{\tau\tau}^{\left(\alpha\right)},g_{\tau\tau}^{\left(\alpha^2\right)},g_{\rho\tau}^{\left(\alpha^2\right)}, g_{i\tau}^{\left(\alpha^2\right)}$ and higher order contributions since these do not affect the Einstein equation to the order that we consider.}
\begin{align}\label{timenonlin}
&\frac{ds^2}{\ell^2} = -\frac{\rho}{\alpha}d\tau^2 \\&\nonumber+\rho^2 d\tau^2 +  2 d\tau d\rho +d\Omega_2^2
+\left(1 - {\rho} \right) \left[ v^2 d\tau^2 - 2 v_i d\tau dx^i \right]-2\rho P d\tau^2\\&\nonumber
+\alpha \Bigl[\left(-4\rho+ 2 P\right) d\Omega_2^2+(1-\rho)  {v_i v_j} dx^i dx^j\\&\nonumber-{\left(\rho^2-1\right)} \left(\nabla^2 v_i+
  R_{j i} v^j \right) d\tau dx^i -2 v_i d\rho dx^i + \left(v^2 + 2 P\right){d\tau d\rho}  + 2    \left(1 - {\rho}\right) \phi_{i}^{(\alpha)} d\tau dx^i\Bigr]  \\&\nonumber
 +\alpha^2 \left(4\rho^2d\Omega_2^2+2 g_{\rho i}^{(\alpha^2)}  dx^id\rho + g^{(\alpha^2)}_{ij} dx^idx^j
  \right) + \dots.
\end{align}
The $v^i$, $P$  and $\phi_i^{\left(\alpha\right)}$ are functions of $(\tau,\Omega^i)$ only
while the $g_{i \rho}^{\left(\alpha^2\right)}$ and $g_{ij}^{\left(\alpha^2\right)}$
are functions of $(\tau,\rho,\Omega^i)$. We have chosen a gauge where $g_{\rho\rho}=0$.
As boundary conditions we require the perturbations to preserve the induced metric on the hypersurface $\rho=1$
\beq
ds_{3d}^2= \left(-\frac{1}{\alpha} + 1\right)d\tau^2 + \left(1-2 \alpha\right)^2 d\Omega^2_2~,
\eeq
 up to a conformal factor
\beq
1 + 2 \alpha P + \mathcal{O}(\alpha^2)~.
\eeq
We also study perturbations such that this hypersurface has constant mean extrinsic curvature and that the solution is regular at the future horizon $\rho=0$. These boundary conditions are the natural extension of the boundary conditions imposed on the linearized fluid modes of the last section.

We now examine the conditions on the deformation parameters imposed by the Einstein equation with a positive cosmological constant ${\cal G}_{\mu\nu}=0$ up to and including ${\cal O}(\alpha^0)$. We further assume that the only excited field is the metric. The first non-trivial condition appears at ${\cal O}(\alpha^{-1})$. Here, for ${\cal G}_{\tau\tau}=0$ to be satisfied, the velocity field $v^i$ is required to be incompressible. At the next order ${\cal O}(\alpha^0)$, the non-trivial equations are ${\cal G}_{\tau\tau}={\cal G}_{\tau i}=0$. From the ${\cal G}_{\tau i}=0$, it follows that the $(v^i,P)$ need to satisfy the non-linear incompressible Navier-Stokes equation (\ref{navierstokesfluid}) on a unit $S^2$. Our result is in complete accordance with the linearized analysis.\footnote{We would not expect to see the pressure in the linearized analysis since at the linear level vector and scalar representations of $SO(3)$ decouple. This is no longer the case at second order in perturbation theory where we expect the equation for the vector representation to be affected by scalars as in the non-linear case.
As in the linearized analysis, we have not considered sound modes, which would contribute to the divergence of $v^i$. 
}

From the ${\cal G}_{\tau\tau}=0$ Einstein equation at $\mathcal{O}(\alpha^0)$ we find the requirement
\beq\label{phie}
\nabla_{S^2}^i \phi_{i}^{(\alpha)} =2\p_\tau P + \p_\tau (v^2) +\textbf{total derivatives on the 2-sphere}.
\eeq
The above relation implies:
\begin{equation}
\partial_\tau \left(  \int v^2 d\Omega_2 \right) = -2 \partial_\tau \left(  \int P d\Omega_2 \right)~.
\end{equation}
There is a similar condition between the velocity and pressure fields in the Schwarzschild case \cite{Bredberg:2011xw}.\footnote{It has been speculated that such an integral relation is related to changes in the horizon area~\cite{Bredberg:2011xw, E09}. } 
This component of the Einstein equation also determines a scalar function involving $g_{\rho i}^{(\alpha^2)},g_{i j}^{(\alpha^2)}$. 

\subsection{Deformations of the Fluid}

A natural question to ask about the fluid is whether one can deform it. In this section, we discuss two simple examples of possible deformations of the fluid.

The first is given by adding a small non-rotating black hole of mass $M$  at the origin of the static patch. This changes the $-g_{tt} = g^{rr}$ components of the metric (\ref{static}) to $V(r) = 1 - (r/\ell)^2 - 2M/r$. For positive values of $M$, adding the black hole has the effect of pulling in the cosmological horizon and thus decreasing its size. For small $\varepsilon \equiv M/\ell$, the new position of the cosmological horizon is given by $r_{cos} = \ell(1 - \varepsilon)$ to leading order. In the analogous case of the Schwarzschild black hole, placing a mass at the center of the static patch corresponds to extracting some mass from the black hole, thus shrinking its horizon. The mass deformation we have described preserves the spherical symmetry of the background and thus the near horizon dynamics are expected to be governed by the Navier-Stokes equation on a sphere.

A slightly more involved deformation corresponds to placing a small rotating mass at the origin of the static patch. This will cause the cosmological horizon itself to rotate. The function determining the positions of the horizons is now given by:
\begin{equation}
V(r)   = (1 + (a/r)^2)\left(  1 - (r/\ell)^2 \right) - 2 M/ r~.
\end{equation}
As with the mass term, adding angular momentum shrinks the cosmological horizon. To lowest order in small $\varepsilon$ and small $\upsilon \equiv (a/\ell)^2$ we find that:
\begin{equation}
r_{cos} = \ell \left(  1 - \varepsilon (1- \upsilon ) - \frac{3}{2}\varepsilon^2 + \mathcal{O}(\varepsilon^3, \varepsilon^2 \upsilon, \varepsilon \upsilon^2,\upsilon^3)   \right)~.
\end{equation}
The angular momentum of the space-time becomes $\mathcal{Q}_{\partial_\phi} = - a M / \left(1 + (a / \ell)^2 \right)$. It should be noted that a finite deformation with angular momentum will also deform the sphere into a spheroidal surface. Thus we lose spherical symmetry and it might be possible that the near horizon dynamics is no longer governed by the Navier-Stokes equation on the round metric of $S^2$.


\section{Pushing the Timelike Surface}

So far we have analyzed the behavior near the cosmological horizon.
Another timelike surface of interest in the static patch is given by the observer's worldline at $r=0$.\footnote{Due to its resemblance with the boundary of AdS in the presence of an eternal black hole, recent work has emphasized the potential importance of the worldline as a candidate for the ultraviolet (holographic) description of the static patch \cite{Alishahiha:2004md,Anninos:2011af}. Although such a holographic duality is far from clear, one expects that the infrared behavior must give rise to the Navier-Stokes equation described in the former section.}
Returning to the analysis of linearized gravity, if we impose Dirichlet boundary conditions leaving the worldline unperturbed for purely outgoing modes, we obtain another set of discrete modes known as quasinormal modes (see for example \cite{LopezOrtega:2006my}). In the original static patch coordinates (\ref{static}) these are given (for the vector modes) by:
\begin{equation}\label{qnm}
\omega_n \ell = - i \left( n + l + 1  \right)~, \quad n = 0, 1, 2, \ldots
\end{equation}
Due to the fact that $l \ge 1$,  a gapless mode is absent in the above spectrum of quasinormal modes. This is in contrast to the fluid modes (\ref{fluid}) which have $\omega_f = 0$ at $l = 1$. The gapless mode is absent due to the fact that the $\omega = 0$, $l = 1$ perturbation diverges on the worldline, as shown in appendix \ref{smallrot}. It reappears in the spectrum as soon as we `puff up' the thickness of the worldline.


\subsection{`Flowing' the Dispersion Relations}

Our aim is to study the behavior of perturbative data on constant $r$ surfaces as we push them from the horizon toward the worldline. There is a clear distinction between the lowest $n=0$ quasinormal modes (\ref{qnm}) and the fluid modes (\ref{fluid}). Given a constant $r$ slice at some position $r = r_c$ we impose Dirichlet boundary conditions leaving the induced metric on the $r=r_c$ unchanged. This constant $r$ surface is directly analogous to the timelike hypersurface at $\rho=1$ considered above. Furthermore, we require that the modes are purely outgoing. As before, these two conditions will only be satisfied for a discrete set of modes, but the dispersion relation will now depend on the dimensionless parameter $x = r_c/\ell$. For the surface near the horizon we have $x \to 1$ and as we approach the worldline we have $x \to 0$. For general $x$, the problem cannot be approached analytically and we must resort to numerics.

The dispersion relation corresponds to the pole structure of the Green's function of the vector modes on the particular cutoff surface. Thus, naturally, a flow of the dispersion relation corresponds to a flow of the Green's function itself.
For an incompressible fluid on an $S^2$ described by (\ref{navierstokesfluid}) we can readily obtain the tree level retarded Green's function of $v^i$ (see for example \cite{Forster:1977zz}).

To perform the analysis, it is in fact more convenient to use the $(\tau,\rho)$-coordinate system introduced in (\ref{nearcosf}). To study different timelike hypersurfaces we fix $\rho = 1$ and tune $\alpha$ from the horizon at $0$ to the worldline at $1/2$. We must then study for what values of (complex) $\omega$ the purely outgoing solutions $\delta g^{out}_{i\tau}$ in~(\ref{metricII}) vanish at the $\rho = 1$ hypersurface. It is relatively straight forward to compute the corrections to the dispersion relation perturbatively in $\alpha$. For instance, to linear order in $\alpha$ we find:\footnote{It is amusing to note that such a correction could be obtained by adding a suitable forcing term to our incompressible Navier-Stokes equation \cite{Forster:1977zz}.}
\begin{equation}\label{viscosity}
\nu = 1 + \frac{\alpha}{2} \left( 5 + 3 k_V^2 \right)~.
\end{equation}
Expression~(\ref{viscosity}) is only reliable for $l^2 \lesssim 2/3\alpha$.

\subsection{Numerical Results}
\begin{figure}[hb!]
\centering\
$\begin{array}{ccc}
\includegraphics[width=0.24\textwidth]{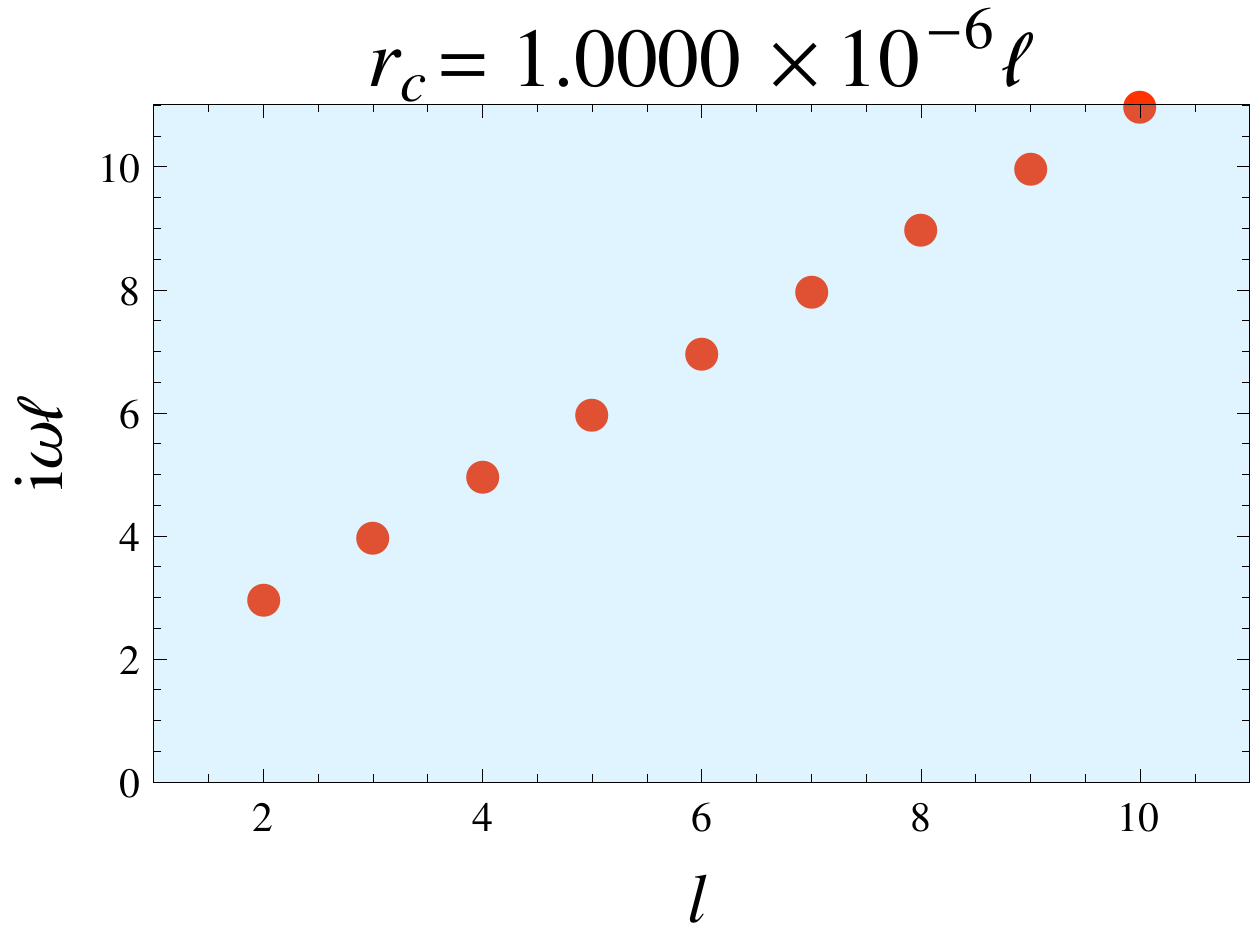}&
\includegraphics[width=0.24\textwidth]{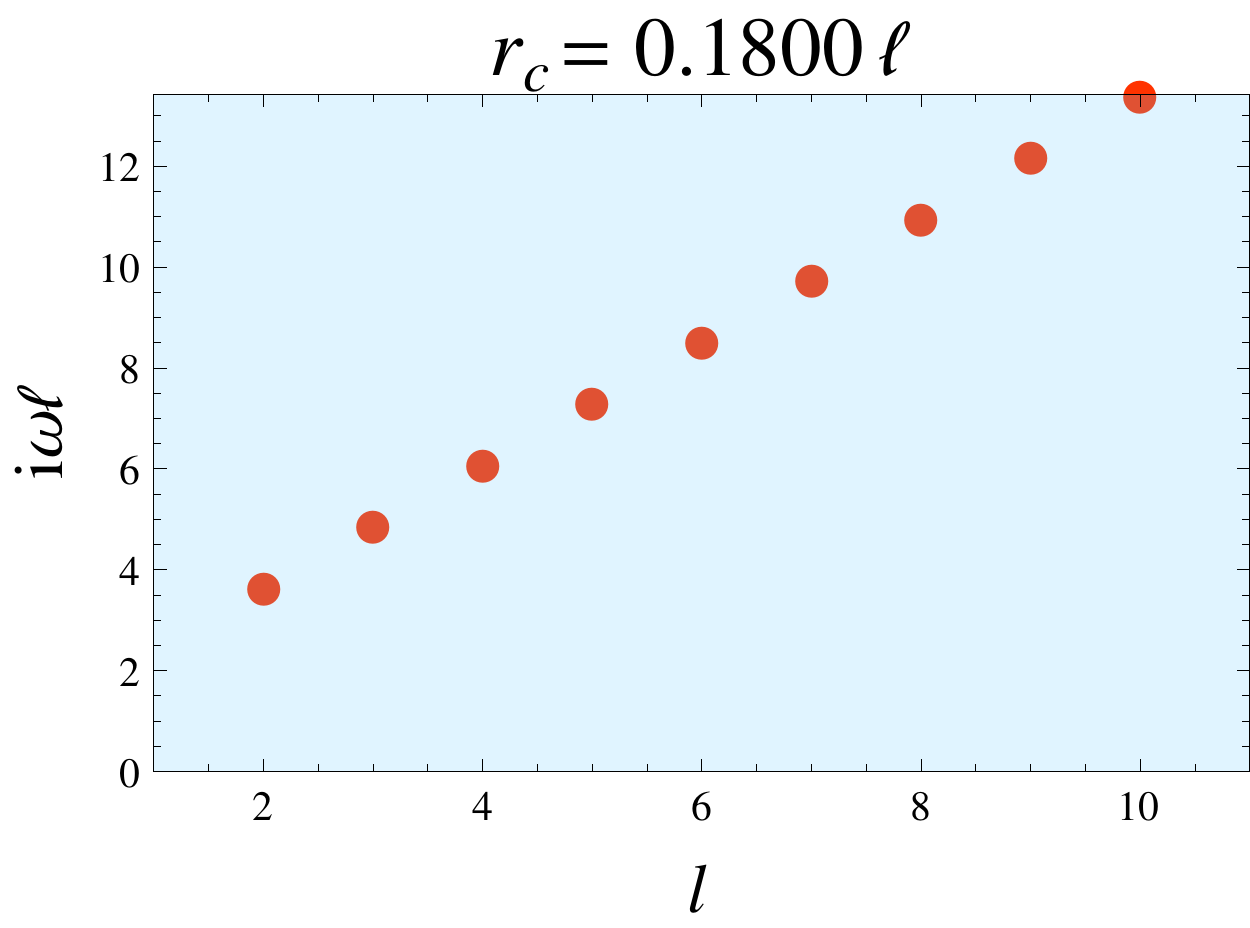}&
\includegraphics[width=0.24\textwidth]{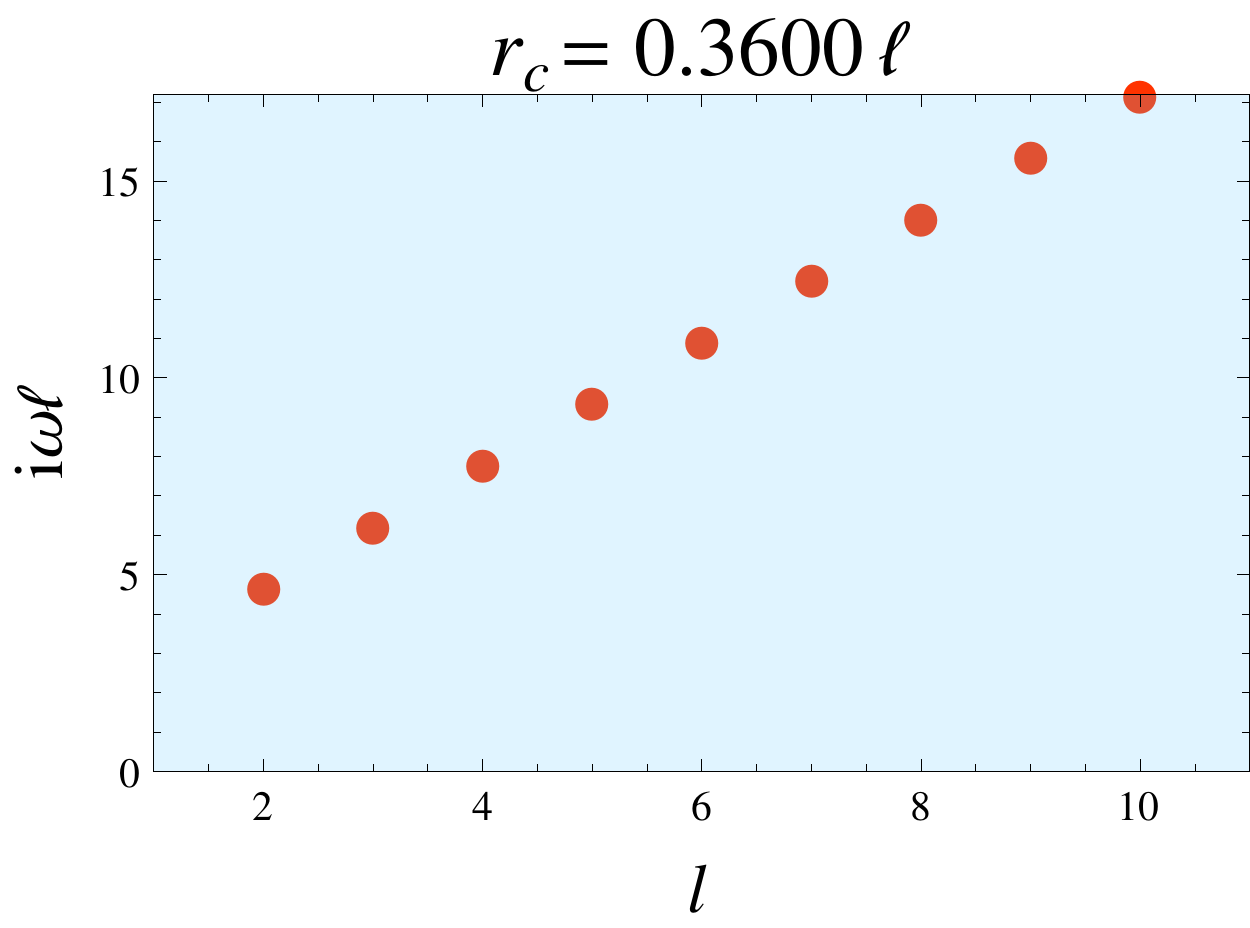}\\
\includegraphics[width=0.24\textwidth]{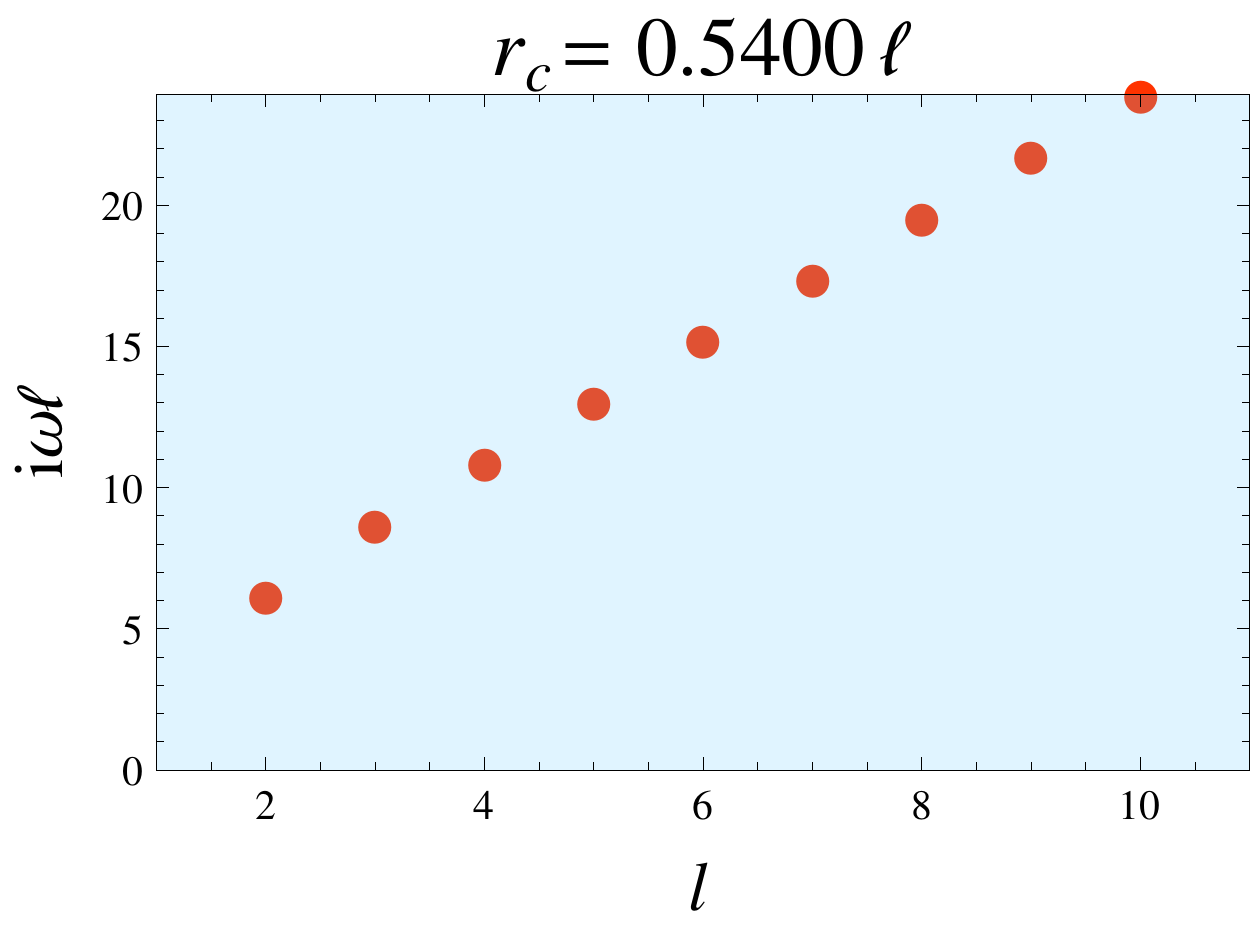}&
\includegraphics[width=0.24\textwidth]{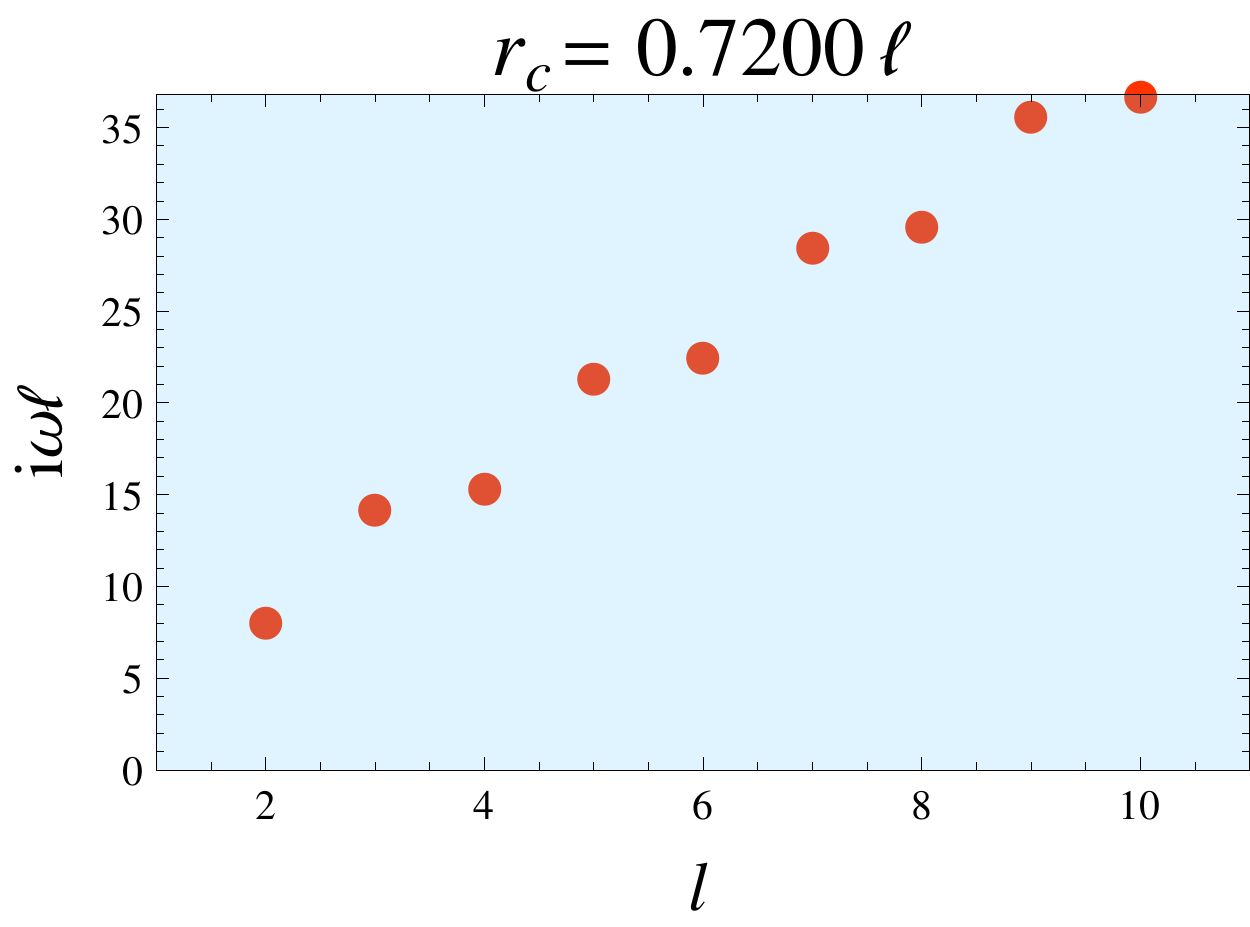}&
\includegraphics[width=0.24\textwidth]{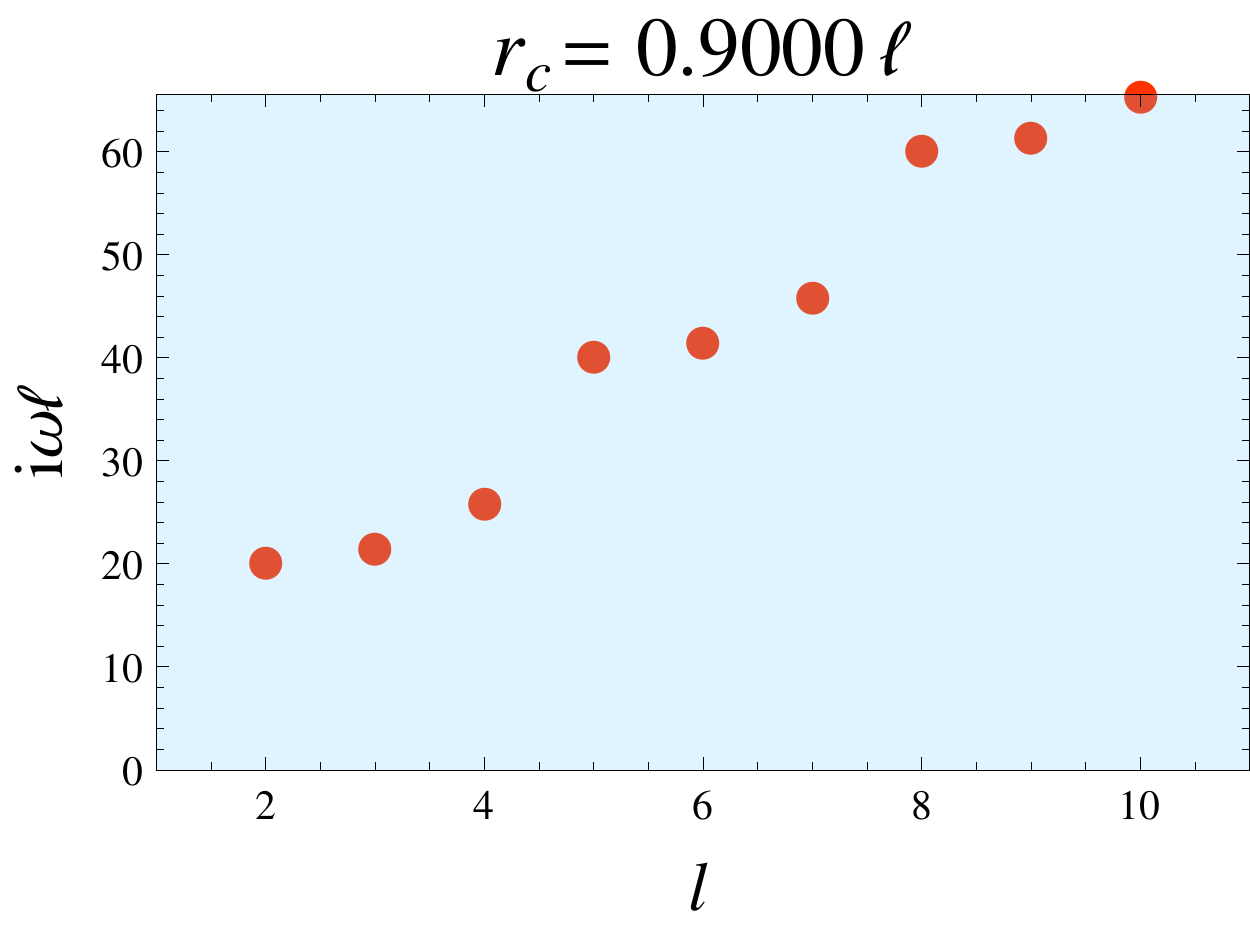}
\end{array}$
\caption{Flow of  frequency spectrum $i\omega \ell$ vs $l$ as we move away from $r_c \approx 0$ toward the cosmological horizon.}\label{flowpoleslinear}
\end{figure}
Since we impose Dirichlet boundary conditions at $\rho=1$, $\alpha$ parametrizes the location of our cutoff surface $r_c$ with respect to the cosmological horizon. The relation is given by
\begin{equation}
\alpha=\frac{\ell-r_c}{2\ell}=\frac{1-x}{2}~.
\end{equation}
As we move $r_c$ away from the cosmological horizon, we expect to deviate from our quadratic dispersion relation~(\ref{fluid}). 
 Generically when searching for zeros of $\delta g^{out}_{i\tau}(\rho=1)$ in~(\ref{metricII}) for arbitrary but fixed $r_c$ and $l$, one runs the risk of finding any one of a tower of such zeros (see (\ref{qnm}) for example).
\begin{figure}[ht!]
\centering\
\includegraphics[width=0.32\textwidth]{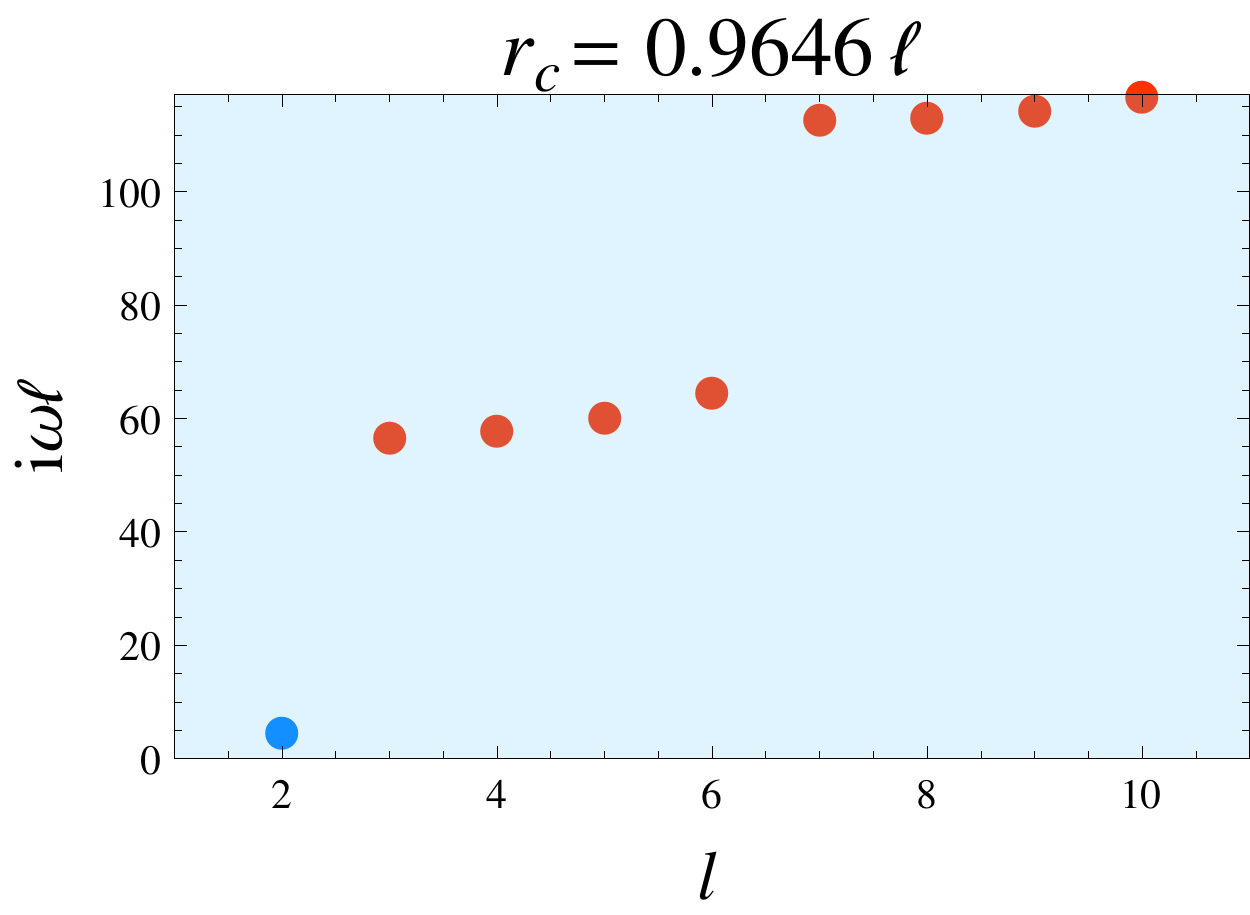}
\includegraphics[width=0.32\textwidth]{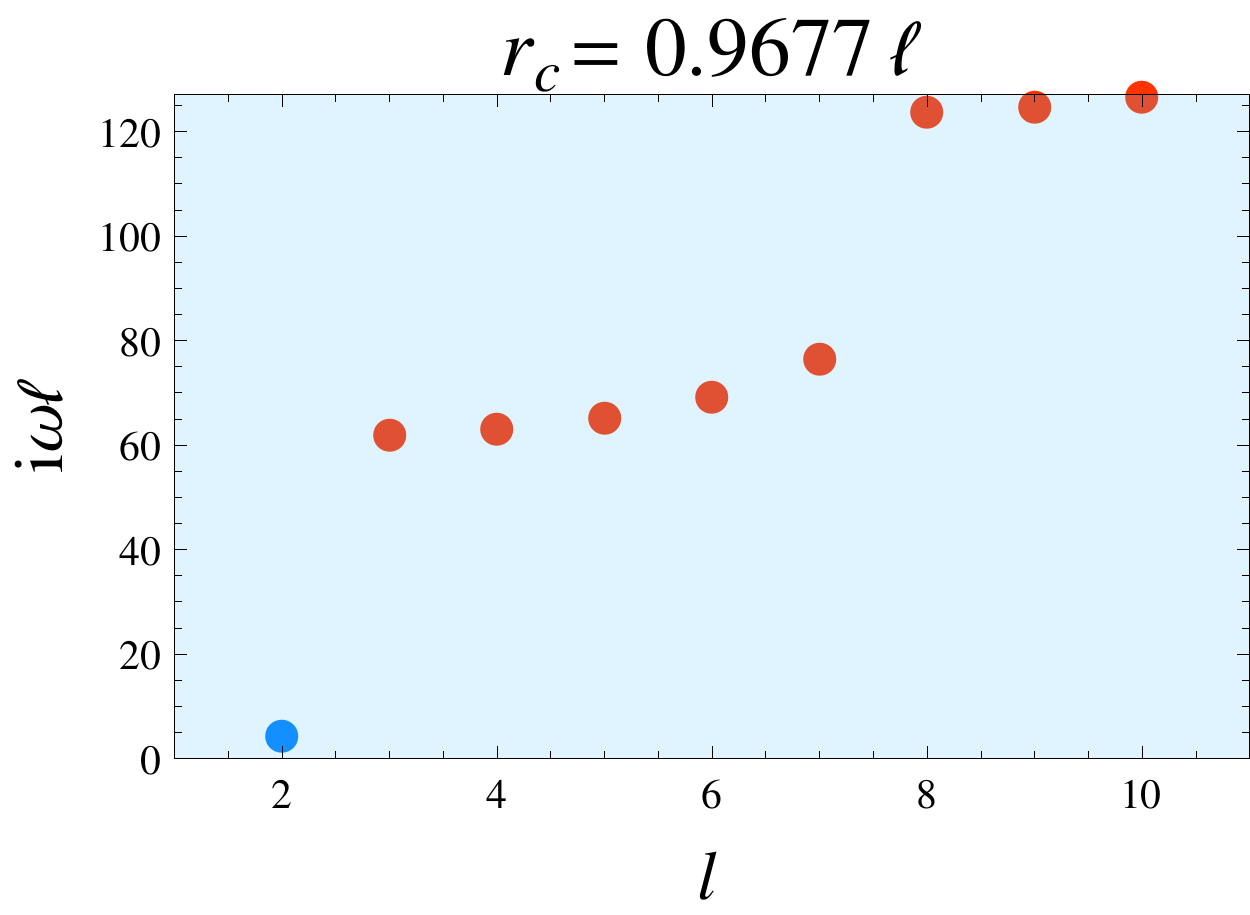}
\includegraphics[width=0.32\textwidth]{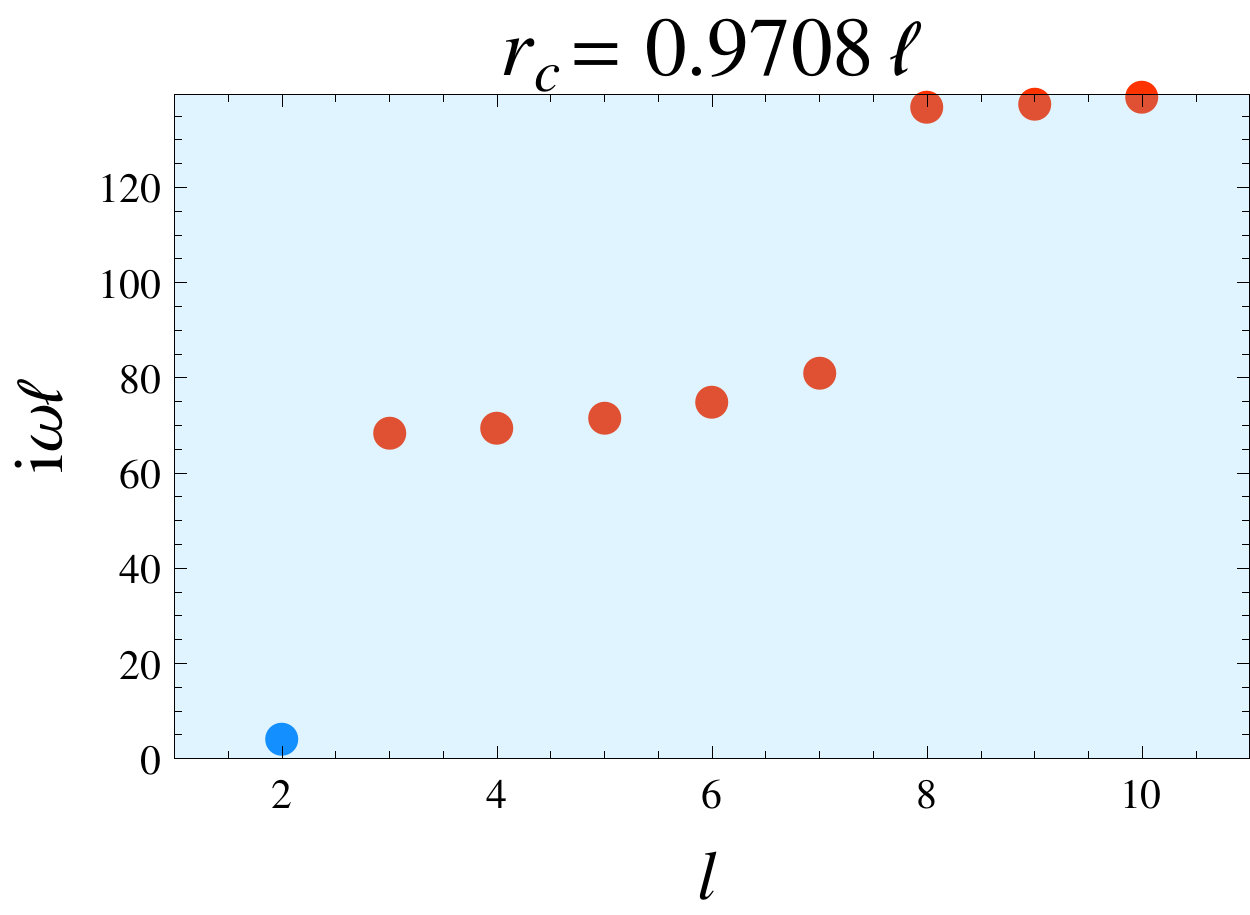}
\includegraphics[width=0.32\textwidth]{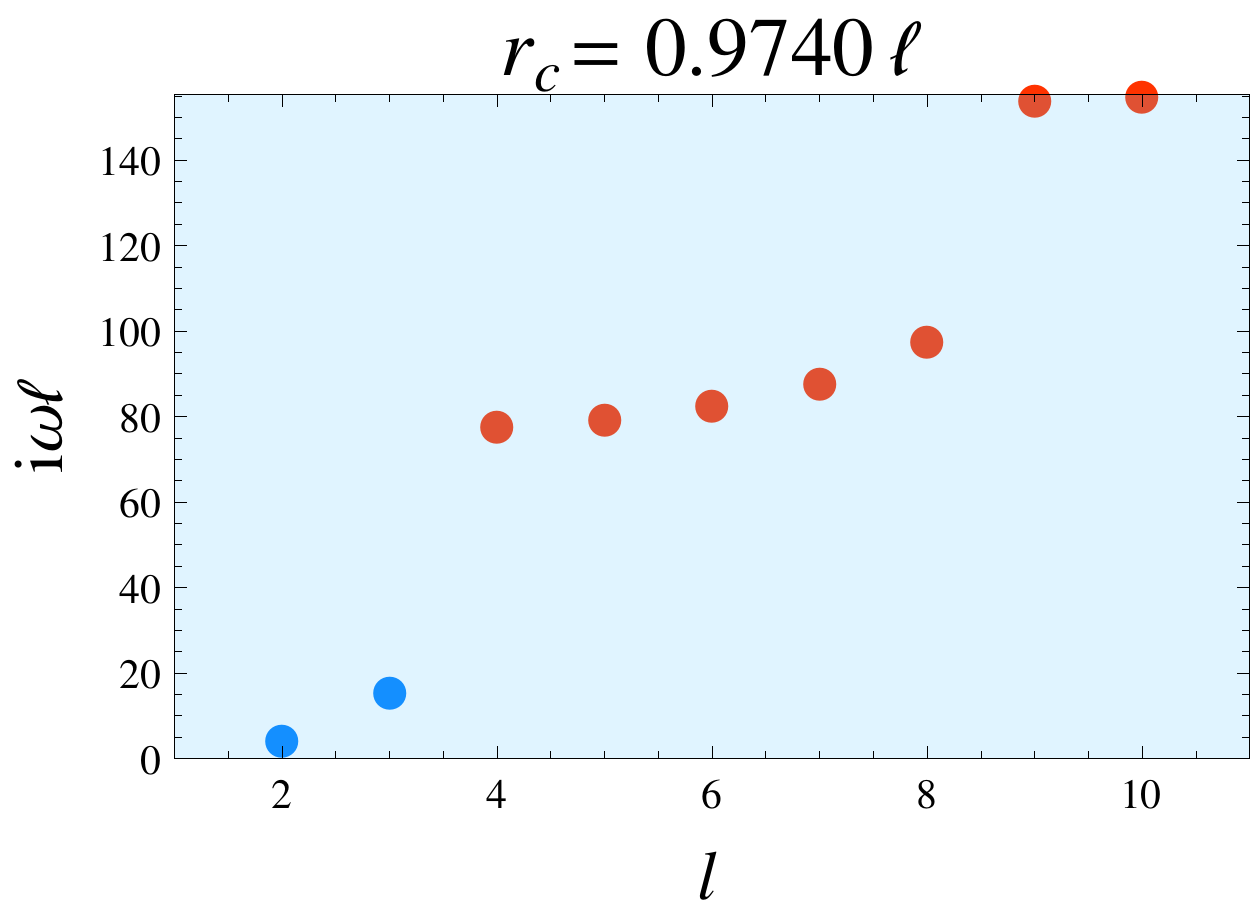}
\includegraphics[width=0.32\textwidth]{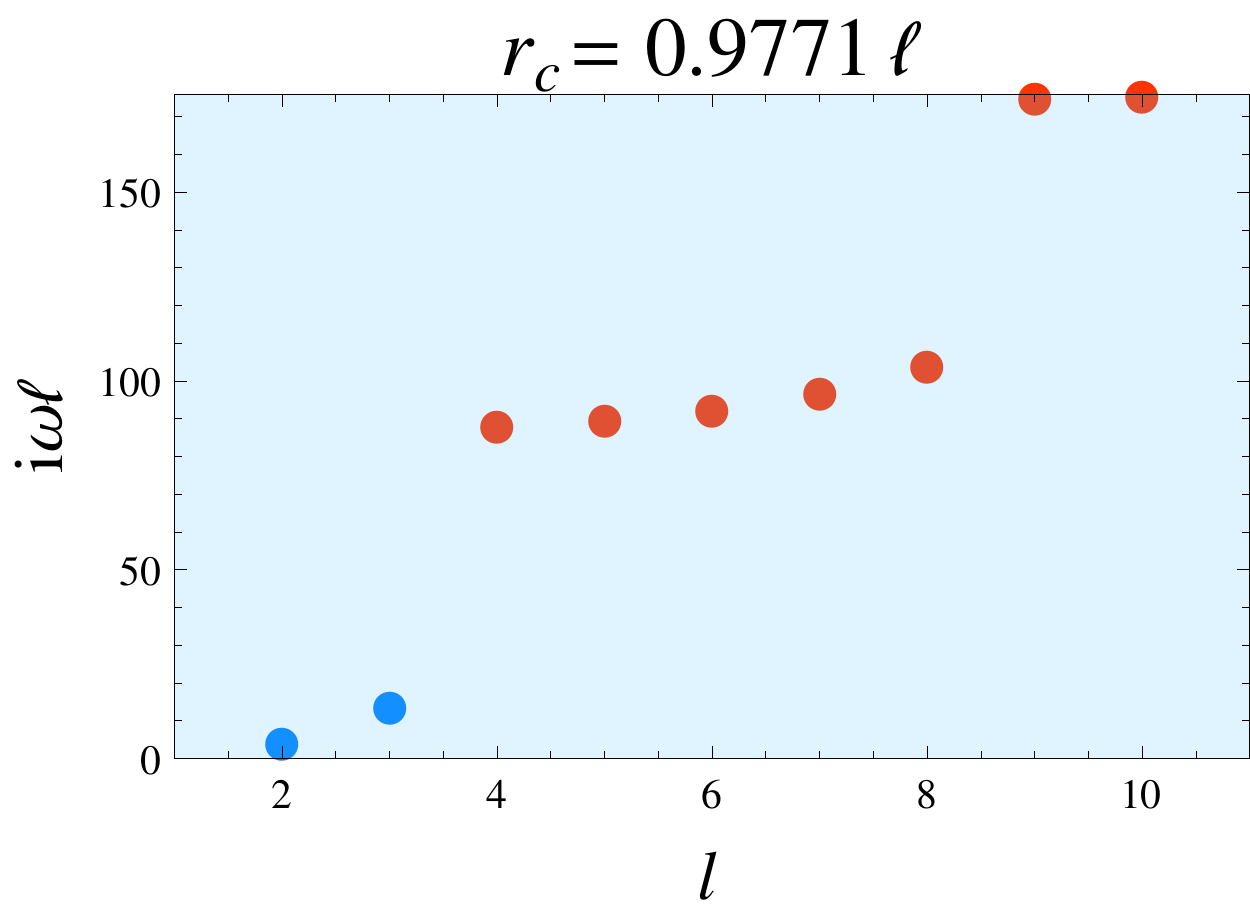}
\includegraphics[width=0.32\textwidth]{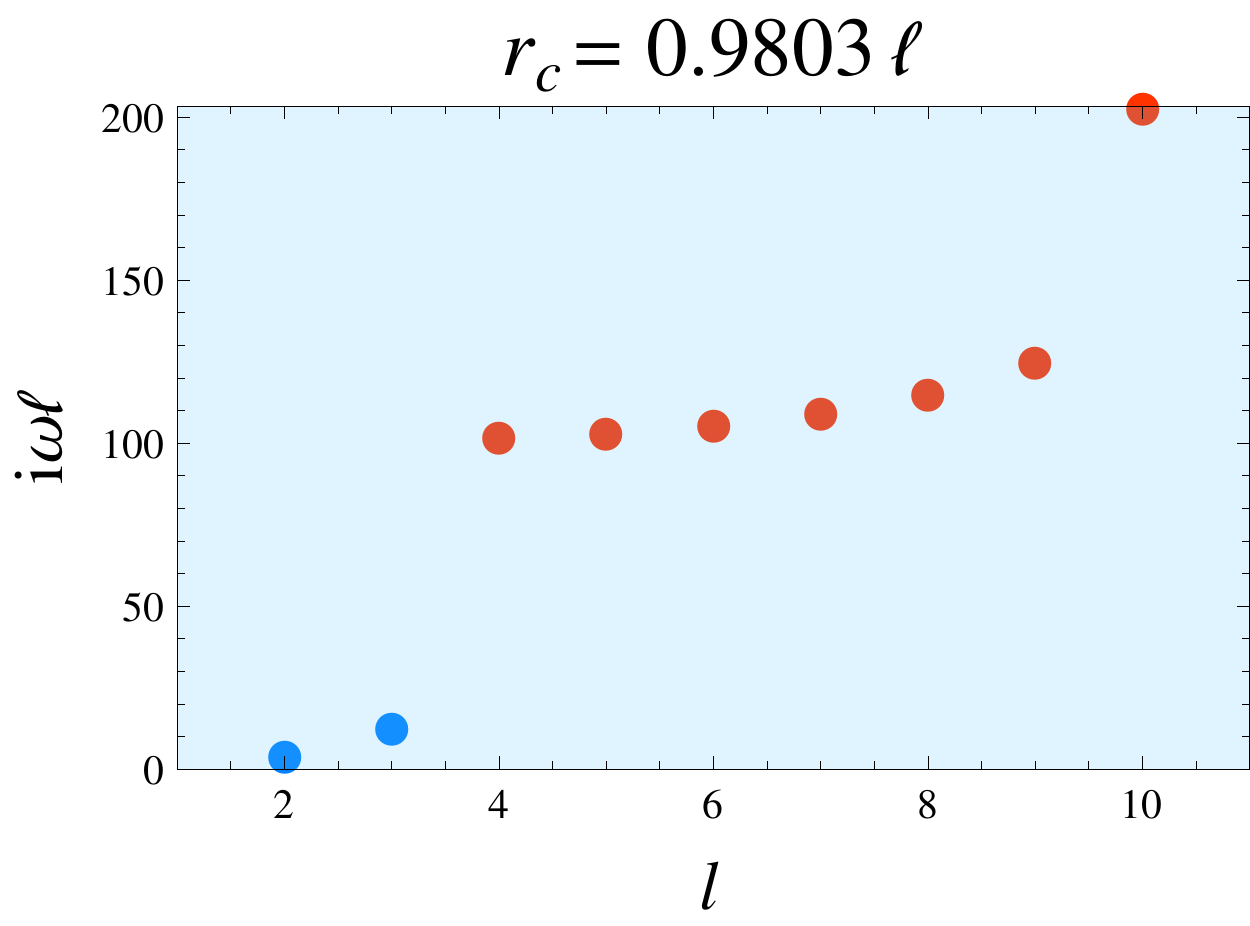}
\includegraphics[width=0.32\textwidth]{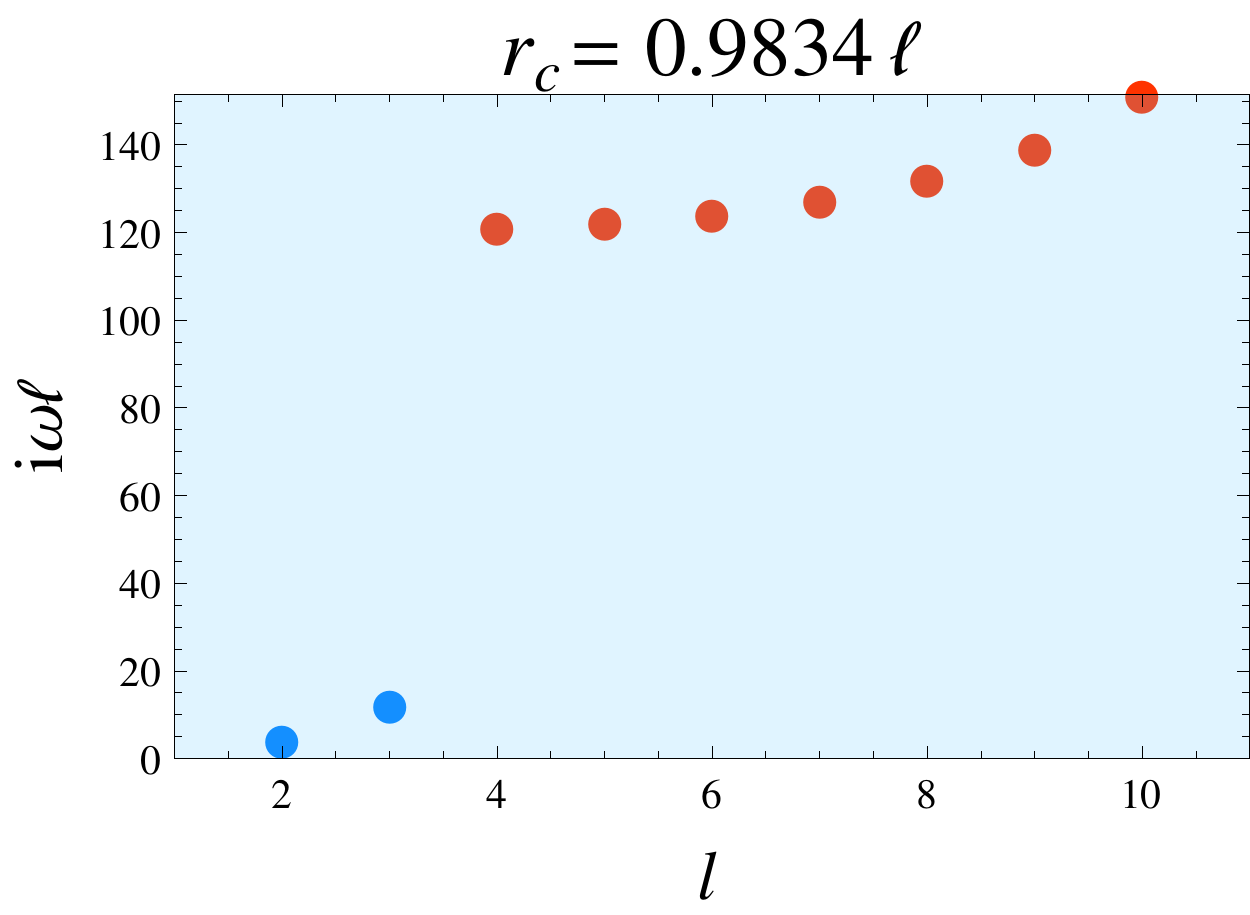}
\includegraphics[width=0.32\textwidth]{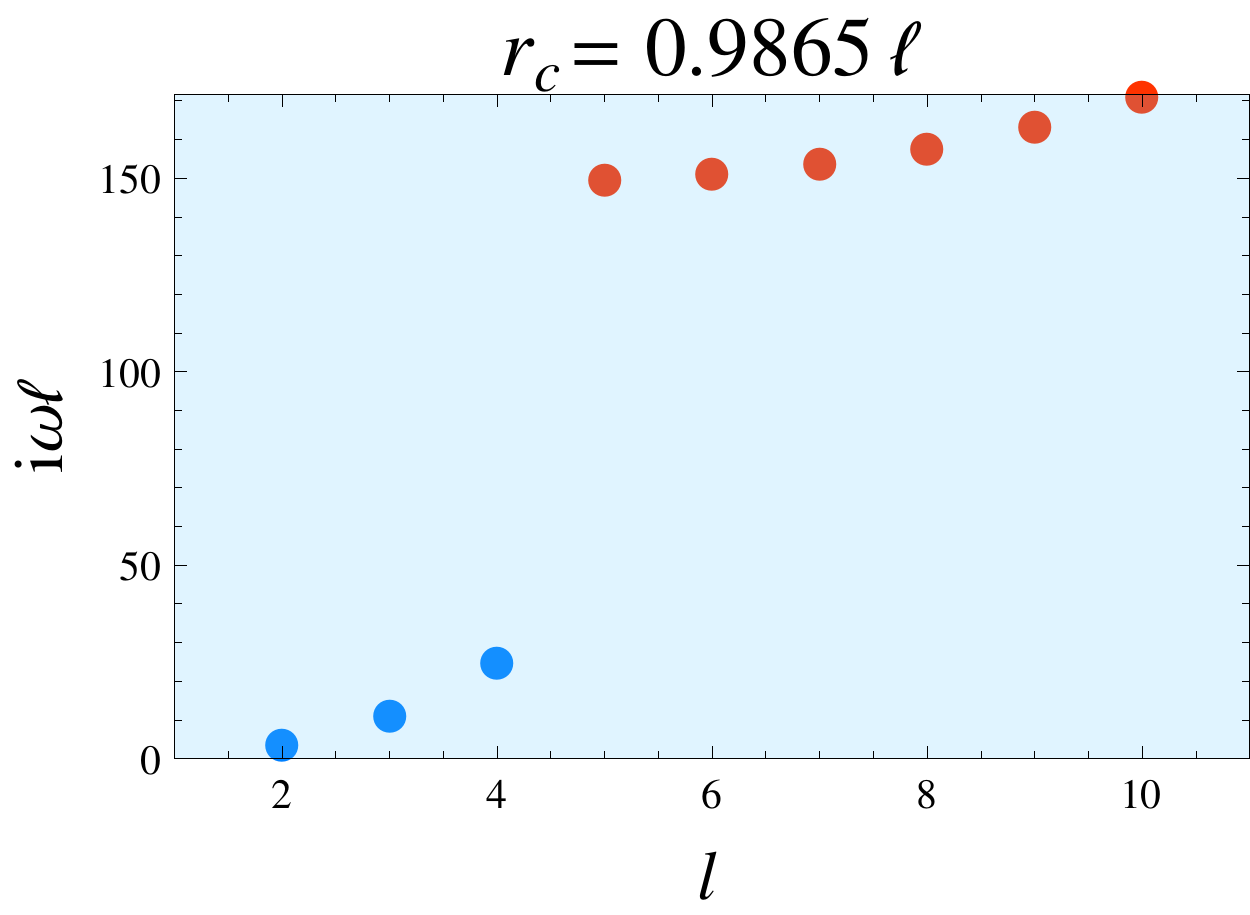}
\includegraphics[width=0.32\textwidth]{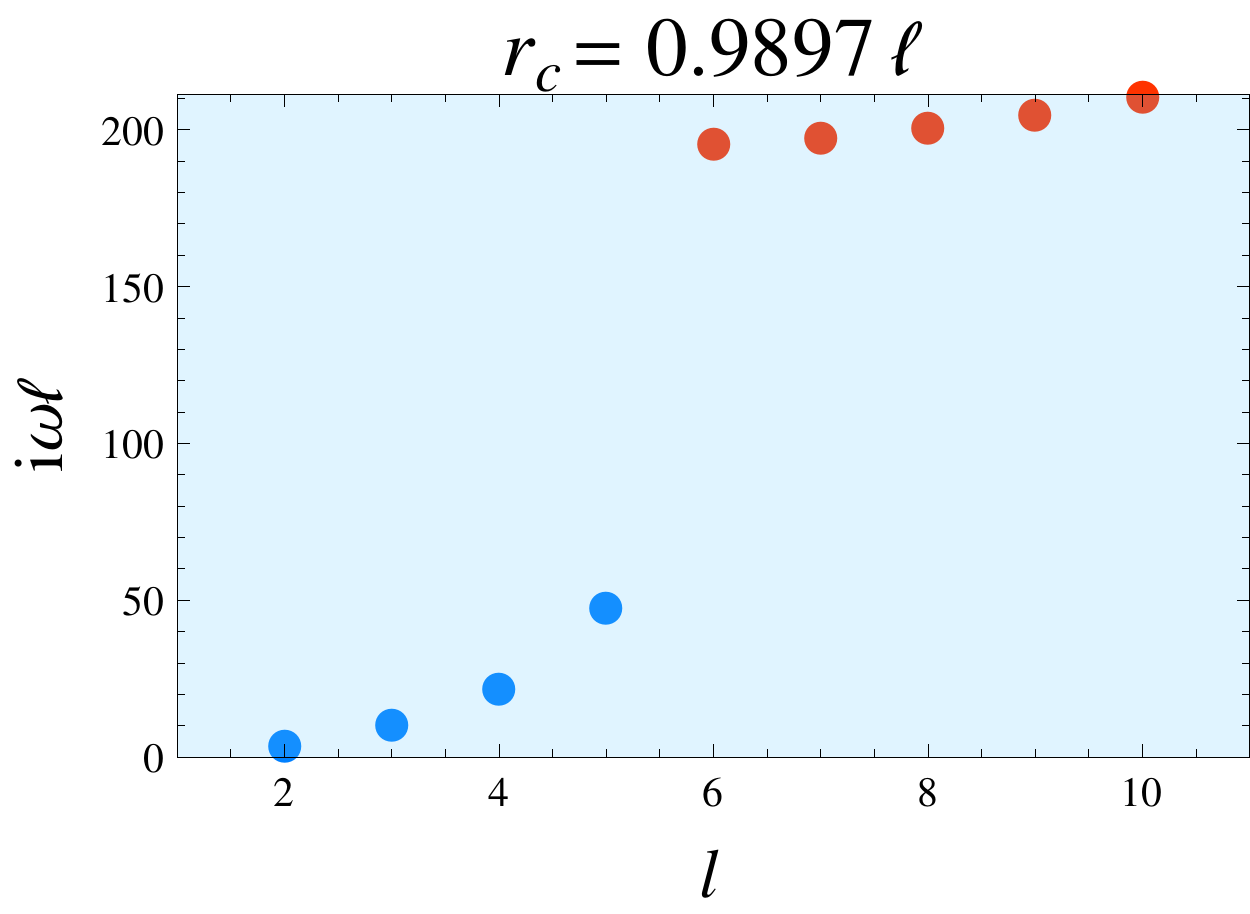}
\includegraphics[width=0.32\textwidth]{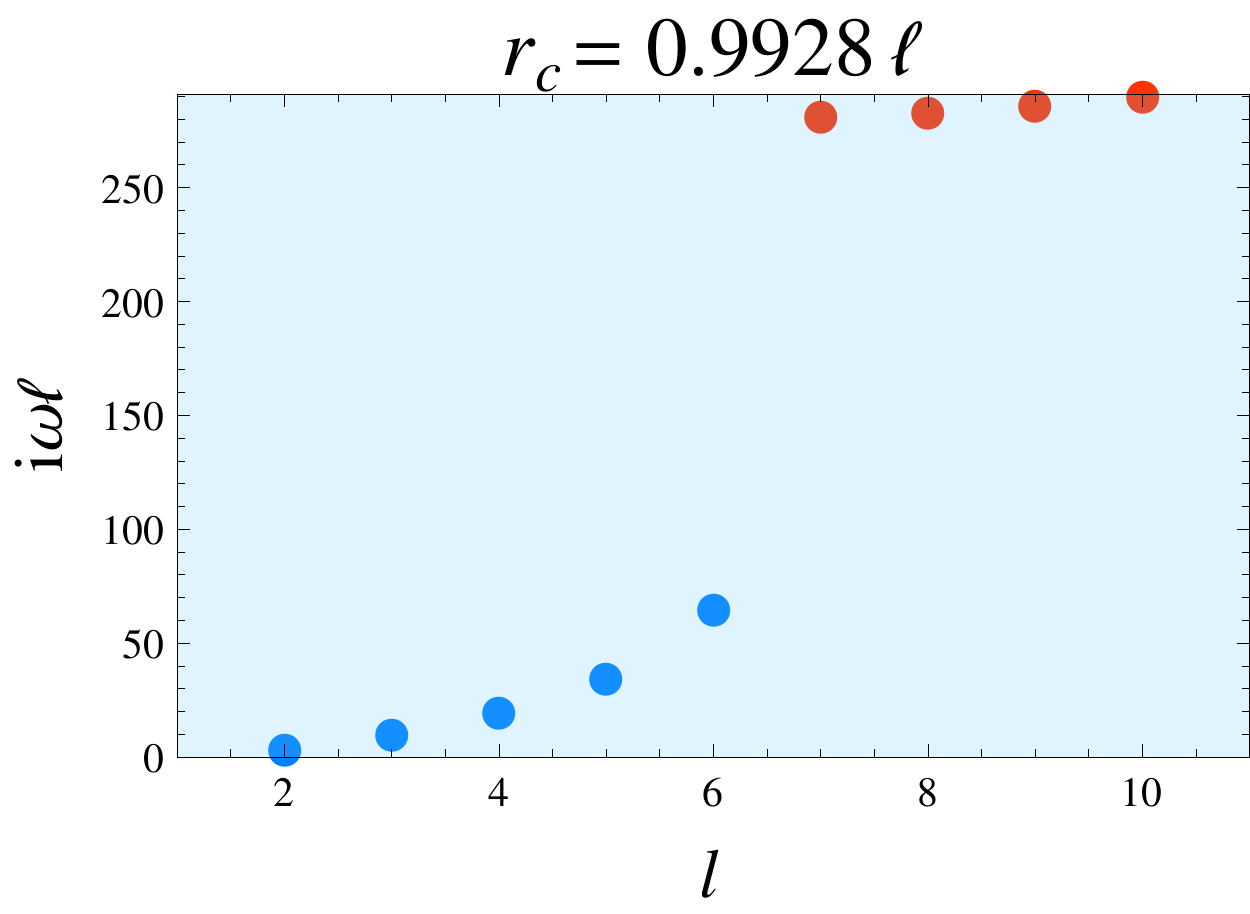}
\includegraphics[width=0.32\textwidth]{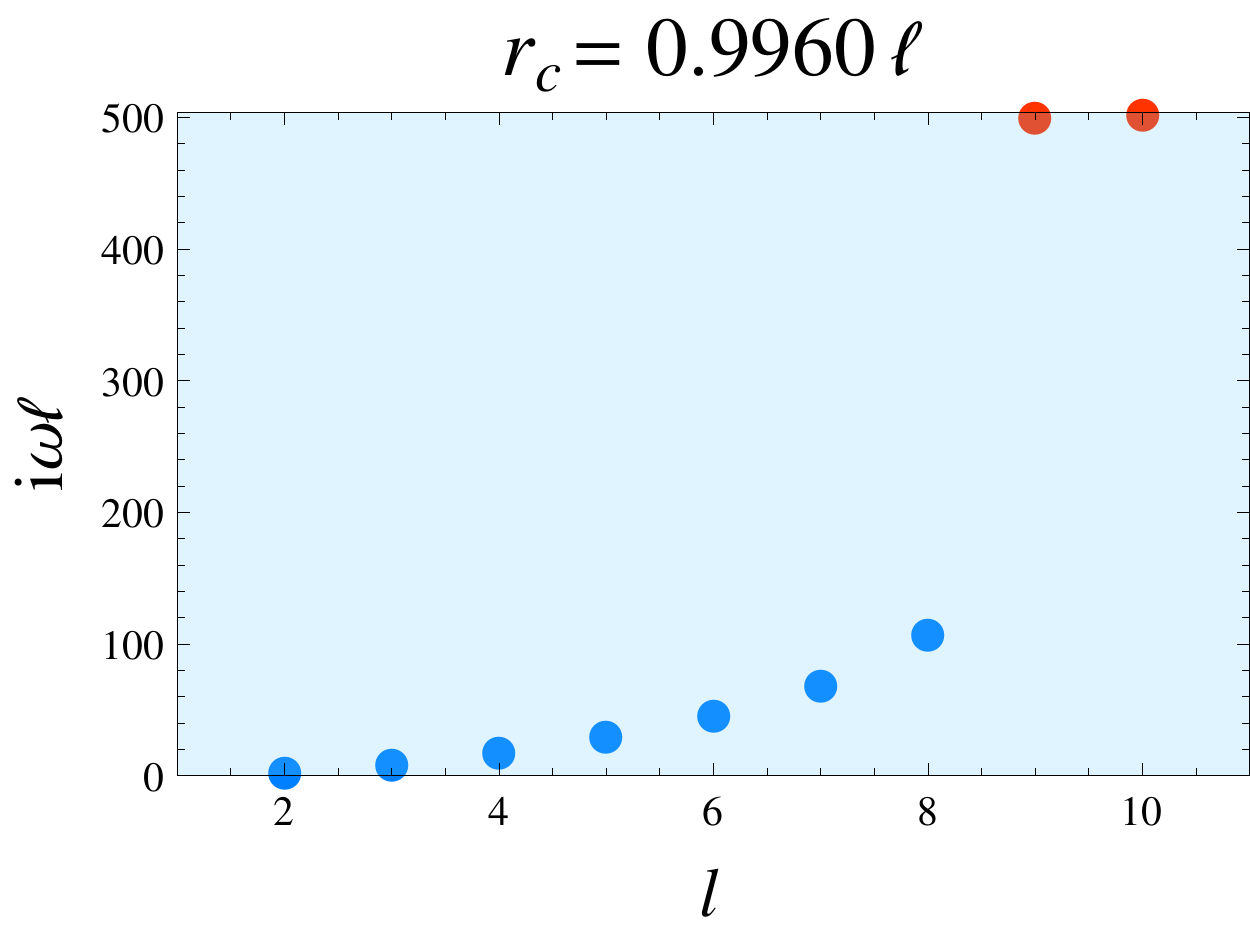}
\includegraphics[width=0.32\textwidth]{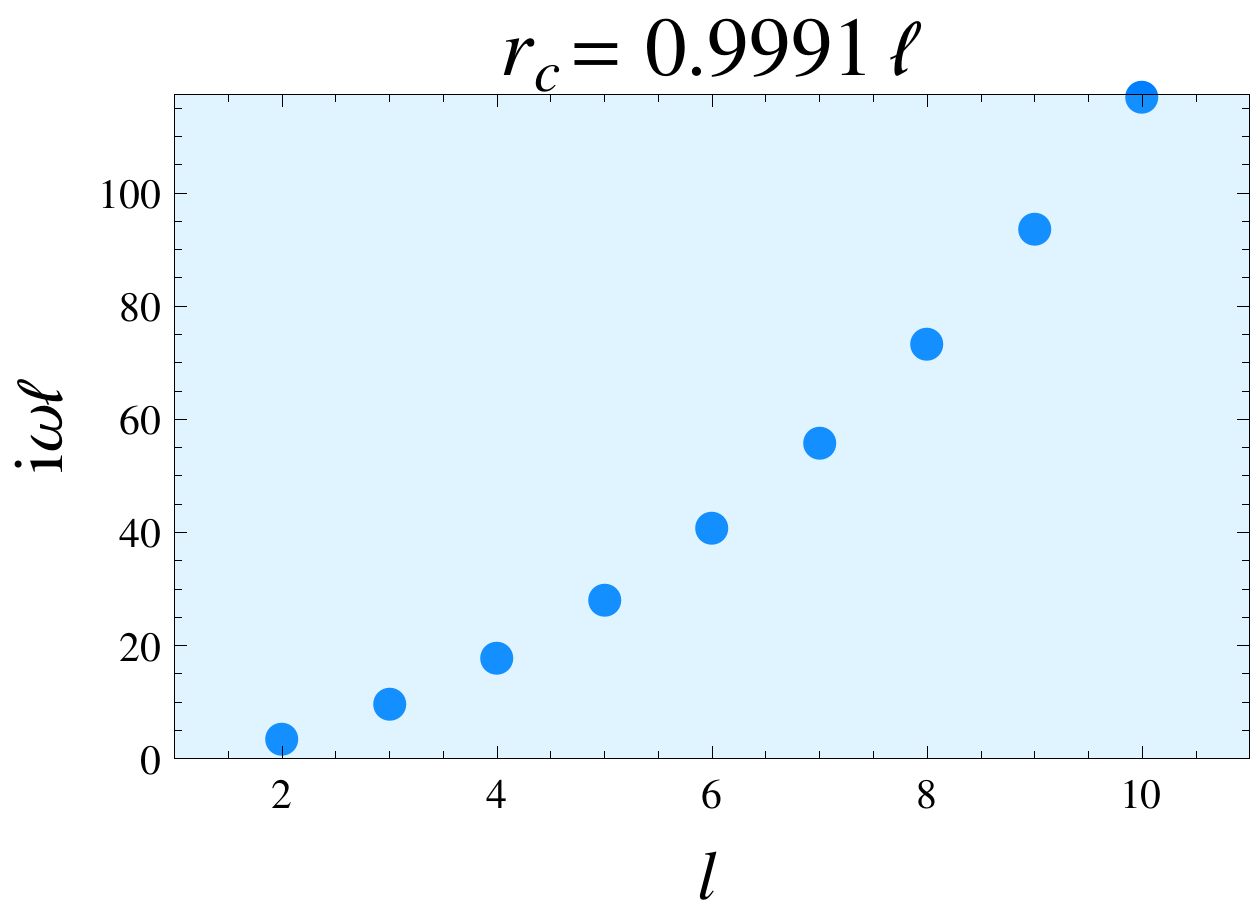}
\caption{Flow of frequency spectrum $i\omega \ell$ vs $l$ as we approach $r_c=\ell$. Points obeying the fluid dispersion relation are represented in blue.}\label{flowpoles}
\end{figure}
\noindent


In order to make our analysis clear, we perform a search for the lowest lying zeros at a given $r_c$ and $l$ and do not present the rest of the tower. In the plots in figures~\ref{flowpoleslinear} and~\ref{flowpoles} we have not searched for the $l=1$ mode as this is where $\omega=0$ and therefore our solutions $\varphi_v^{out}$ and $\varphi_v^{in}$ degenerate.
 We cover this case in appendix~\ref{smallrot}. In what follows we will only examine the case of pure (negative) imaginary $\omega$ given that both the quasinormal modes (\ref{qnm}) and the fluid modes (\ref{fluid}) obey this property. It would be interesting to extend the analysis to general  $\omega$ in the lower complex plane.

We now describe how the spectrum behaves as we approach $r_c=\ell$.
As a reference, we also present the results for $r_c$ close to the pole $r_c=0$ in figure~\ref{flowpoleslinear} where the linear dispersion relation is apparent.
As $r_c$ is increased, we start to observe a deviation from the linear behavior and the poles start
to cluster into staircase-like behavior.
For $r_c$ close to the horizon, starting from the plot at the top left corner of figure~\ref{flowpoles}, we note that there are (at least) three distinct sets of modes separated by large gaps. The $l=2$ mode lies on the fluid dispersion relation line (meaning that it satisfies the dispersion relation given by~(\ref{fluid}) with quartic corrections as in~(\ref{viscosity})), whereas the rest do not. As we move $r_c$ closer to the horizon, we see that the non-fluid modes get pushed higher and in the fourth plot, the $l=3$ mode is plucked down to the fluid line. This happens once again for the $l=4$ mode near $r_c=0.9865\ell$ whereas the non-fluid modes keep getting pulled higher. The reason for these jumps is that the lowest lying zeroes of $\delta g^{out}_{i\tau}$ are modified discontinuously as a function of $r_c$ as is visually depicted in figure \ref{mindfig} of appendix \ref{mindgap}. Finally, we find that arbitrarily close to the horizon the dispersion relation lies entirely on the fluid dispersion relation (\ref{fluid}) computed analytically.

\begin{figure}[ht!]
\centering\
\includegraphics[width=0.4\textwidth]{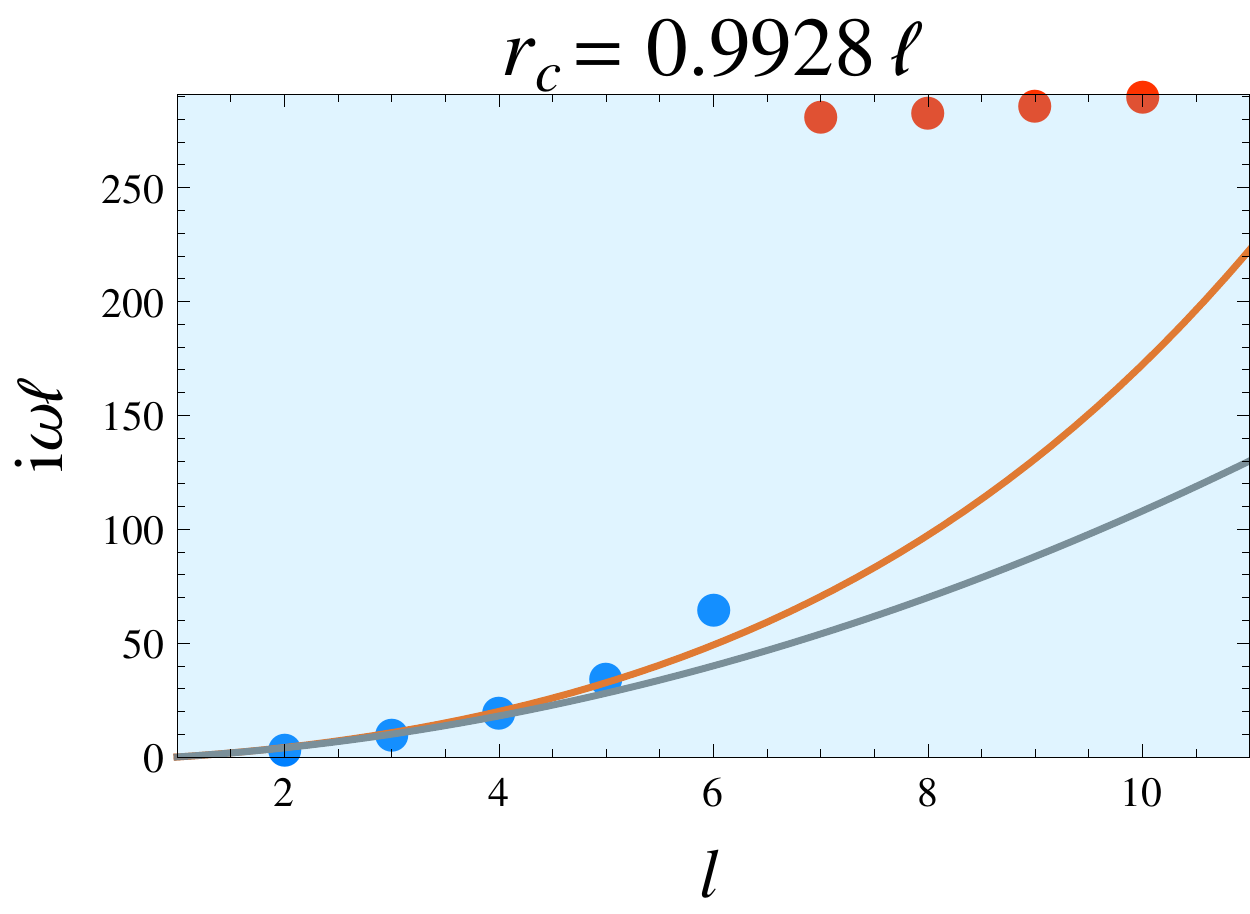}
\includegraphics[width=0.4\textwidth]{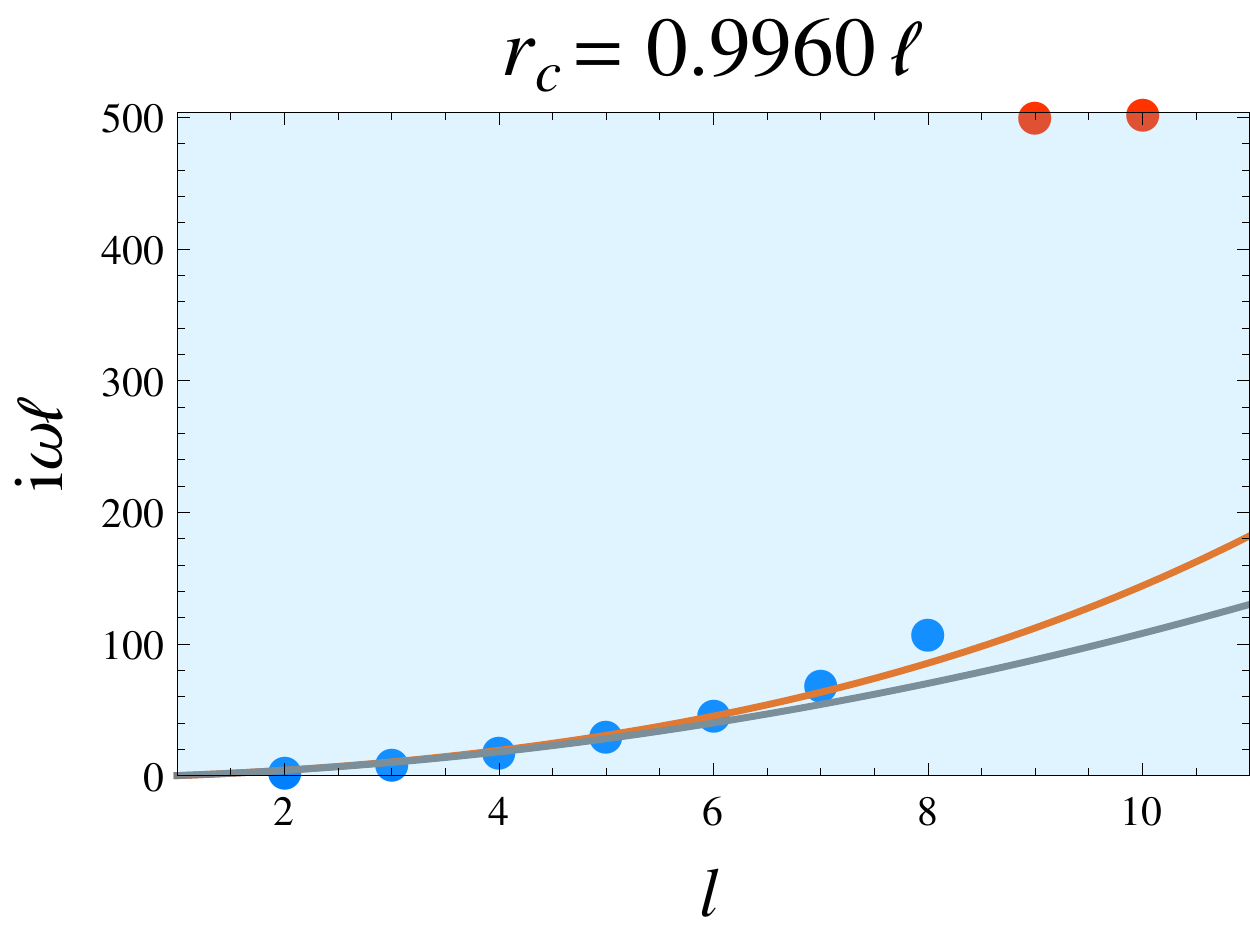}
\includegraphics[width=0.4\textwidth]{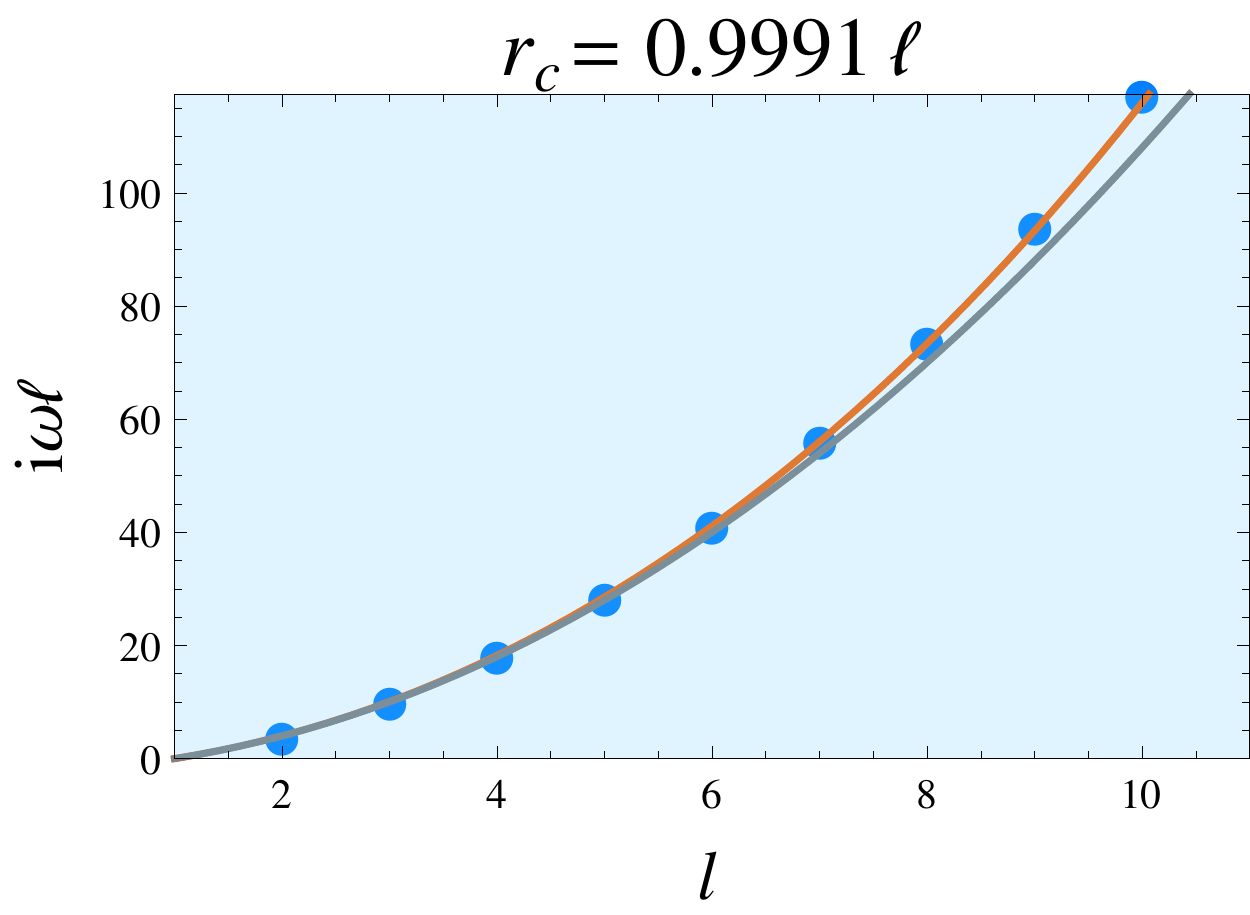}
\includegraphics[width=0.4\textwidth]{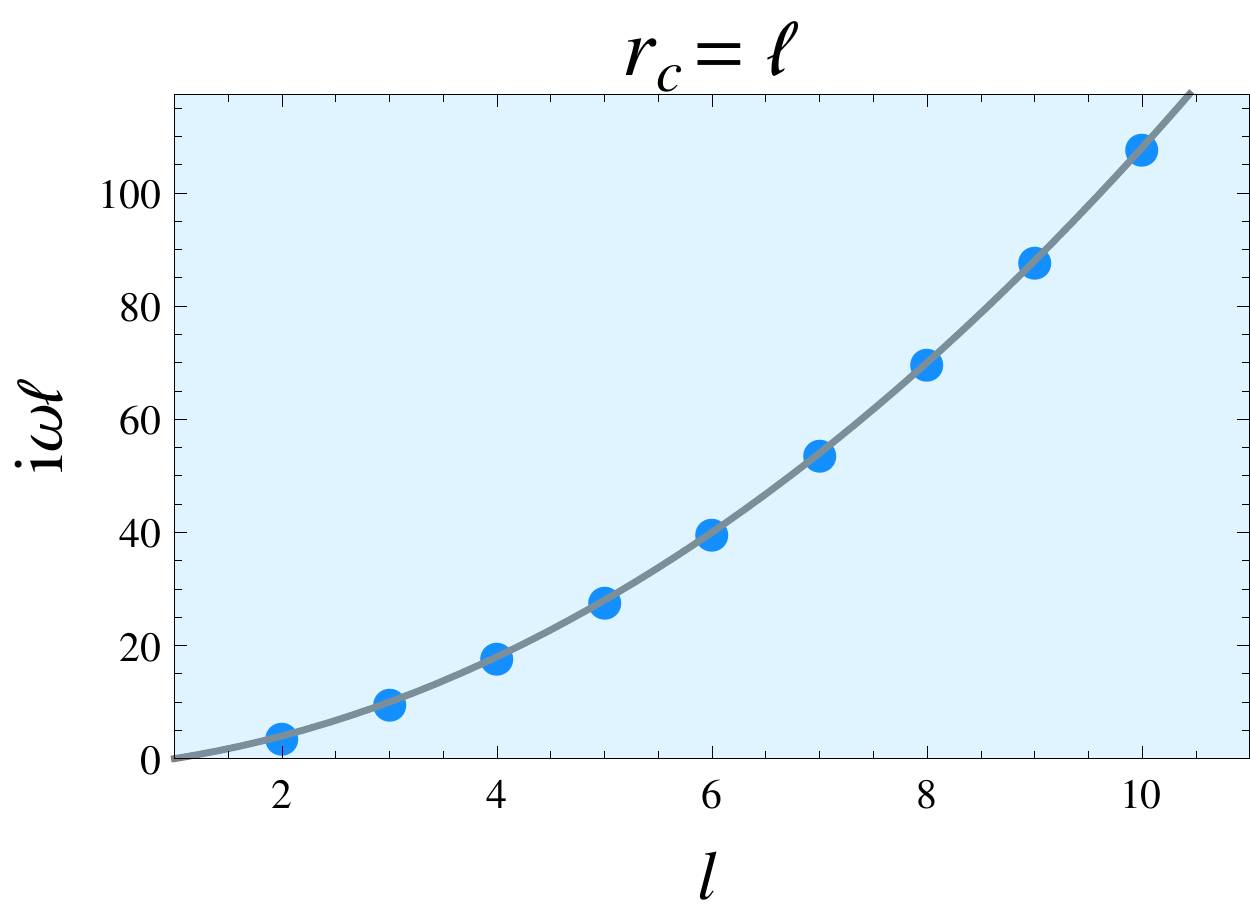}
\caption{Flow of fluid mode frequencies with fit lines. The gray line is given by $i\omega\ell= \left( l(l+1) - 2\right)$ while the orange line is given by $i\omega\ell=\nu(\alpha) \left( l(l+1) - 2\right)$ with $\nu(\alpha)$ given by~(\ref{viscosity}). Note that the orange line fits the data better away from $r_c=\ell$ and the gray and orange lines coincide at the horizon, as expected. }\label{flowpoles2}
\end{figure}
As a check of our analysis in the previous section, in figure~\ref{flowpoles2} we show that the correction at $\mathcal{O}(\alpha)$ of the fluid viscosity gives the correct behavior for the fluid mode frequencies for small but nonzero $\alpha$.
Notice that the orange line in figure~\ref{flowpoles2} fits the data better than the gray line for $r_c$ parametrically displaced from the cosmological horizon. The orange line is precisely the dispersion relation corrected to $\mathcal{O}(\alpha)$ in our analysis of linear perturbations of the background metric of the static patch. The gray line does not include $\mathcal{O}(\alpha)$ corrections.

It is interesting to note that all observed spectra of $i\omega\ell$ are monotonically increasing functions of $l$. This is a feature that we may expect by continuity away from the modes analytically computed at $r_c=\ell$, but holds as we push $r_c$ a finite amount from the surface, even when jumps occur.
%

\section{Incompressible Fluids on Spacelike Slices?}

So far we have discussed several aspects of the static patch, which is the region accessible to a single observer. We would now like to briefly discuss some aspects of the future diamond. After all, future infinity is clearly a viable candidate location for the non-perturbative definition of an asymptotically de Sitter universe. Ordinarily, we would not associate the dual theory on $\mathcal{I}^+$ with the static patch observer. On the other hand, if we impose boundary conditions where there is no incoming flux from the Northern diamond, such that all data reaching $\mathcal{I}^+$ is coming from a single static patch one might conceive of such a relation.\footnote{Such `holographic projections' of the static patch observer were also considered in the rotating Nariai geometry \cite{Booth:1998gf,Anninos:2009yc,Anninos:2010gh}. In that case a near cosmological horizon limit allowed for an isolated space-time whose (spacelike) boundary is of the same type as the spacelike slice we are discussing here.} In this section we will make some simple mathematical observations about the behavior of metric deformations on a spacelike slice just outside the cosmological horizon.

\subsection{Linearized Analysis}

\begin{figure}[ht!]
\centering
\includegraphics[width=0.4\textwidth]{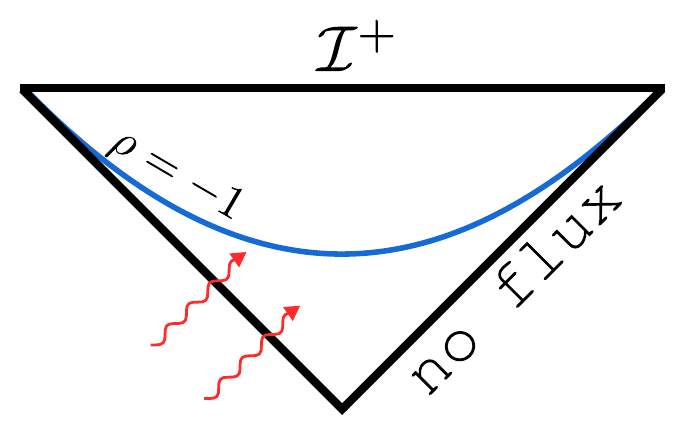}
\caption{Our boundary conditions for the linearized modes are such that the induced metric on the $\rho=-1$ slice is unchanged and there is no incoming flux from the Northern static patch.}\label{futureslice}
\end{figure}

Our aim is to solve the linearized equations in the future diamond, imposing Dirichlet boundary conditions on a constant $r$ surface arbitrarily close to the cosmological horizon. In order to choose an appropriate near horizon coordinate system we begin with the future diamond, described by (\ref{static}) with $r \in [\ell,\infty]$. As before, we introduce the following coordinate transformation:
\begin{equation}
\frac{t}{\ell} = \frac{1}{2\alpha}\tilde{\tau} - \frac{1}{2}\log\left( -  \tilde{\rho} \left(  1 - \alpha \tilde{\rho}  \right)^{-1} \right)~, \quad r = \ell (1- 2\alpha \tilde{\rho} )~.
\end{equation}
In what follows we will drop the tildes. The metric becomes:
\begin{equation}\label{nearcos}
\frac{ds^2}{\ell^2} = \left( -  \frac{\rho}{\alpha} + \rho^2  \right)d\tau^2 + 2d\tau d\rho + \left( 1 -  2\rho \alpha \right)^2 d\Omega_2^2~.
\end{equation}
Taking $\alpha \to 0$ with $\alpha>0$, constant $\rho$ slices for $\rho < 0$ now correspond to spacelike slices just above the cosmological horizon. This slice receives data from the future horizons of \emph{both} the Southern and Northern patches. If we are to isolate one of the observers, say the Southern observer, we must impose that the incoming flux from the Northern static patch vanishes, as shown in figure~\ref{futureslice}.

The vector mode with vanishing flux from the Northern patch is generated by a master field $\Phi^S_v = e^{-2i\alpha\omega t(\tau,\rho)/\ell}\varphi^S_v$, with:
\begin{equation}
\varphi_v^S = (-\rho)^{-i\alpha \omega} {_2}F_1 \left[  a_1,b_1;c_1;  \alpha\rho (-1+2\alpha\rho)^{-1} \right] \; (1 - \alpha \rho)^{ - i\alpha \omega} ( 1 - 2\alpha \rho )^{2i\alpha\omega}~,
\end{equation}
where
\begin{equation}
a_1 = -l - 2i\alpha\omega~, \quad b_1 = 1+l - 2i\alpha \omega~,\quad c_1 = 1-2i\alpha\omega~.
\end{equation}
The linearized metric deformation is given by:
\begin{eqnarray}
\delta g^S_{i \tau} &=& 2 \mathcal{V}_i \times e^{-i\omega\tau} (-\rho)^{i\alpha\omega+1} (1-\alpha\rho)^{-i\alpha\omega+1}\left(  1 - \frac{(1-2\alpha\rho)}{2\alpha}\partial_\rho  \right)\varphi_v^S~,\\
\delta g^S_{i \rho} &=&-2\alpha\mathcal{V}_i \times e^{-i\omega\tau} \;  \left( \frac{1-\alpha\rho}{-\rho}\right)^{-i\alpha\omega}\left[ \left(1-\frac{i\omega(1-2\alpha\rho)}{2\rho(1-\alpha\rho)}\right)- \frac{(1-2\alpha\rho)}{2\alpha} \partial_\rho \right]\varphi_v^{S}~.\nonumber\\
\end{eqnarray}

Demanding that the above vector modes vanish at the spacelike $\rho = -1$ hypersurface in the limit where $\alpha \to 0$ gives the discrete relation:
\begin{equation}\label{sfluid}
\omega_{sf} = + i \left( l(l+1)-2 \right)~,\quad l = 1,2,\ldots
\end{equation}
This is precisely the same dispersion relation that was found earlier in (\ref{fluid}) for the Lorentzian hypersurface but with $\nu_{sf} = - 1$.\footnote{We note that a time-reversal $\tau\rightarrow-\tau$ transformation leads to $v^i \rightarrow -v^i$ and thus introduces a sign change to the viscosity term $\nu_{sf}$ in the non-linear Navier-Stokes equation~(\ref{navierstokesfluid}). Thus, one can interpret a negative viscosity fluid as a time-reversed version of a positive viscosity fluid. We thank R. Loganayagam for pointing this out to us.} To linear order in $\alpha$ we find:
\begin{equation}
\nu_{sf} = -1 + \frac{\alpha}{2} \left( 5 + 3 k_V^2 \right)~.
\end{equation}
It is not hard to show perturbatively in $\alpha$ that $\nu_{sf} (\alpha)= - \nu (-\alpha)$. Upon evaluating the linearized metric modes on the `spacelike fluid' frequencies (\ref{sfluid}) we find that the vector modes are regular as we approach the $\rho \to 0$ horizon.

\subsection{Non-linear Analysis}

We would now like to perform a non-linear analysis of the metric deformations in an $\alpha$ expansion, again in the context of the spacelike slices. The cutoff surface will now be at $\rho=-1$. Non-linear deformations analogous to those presented for timelike slicings are:
\begin{align}
&\frac{ds^2}{\ell^2} = -\frac{\rho}{\y}d\tau^2 \\&\nonumber+\rho^2 d\tau^2 +  2d\tau d\rho + d\Omega^2_2
+\left(1 + {\rho} \right) \left[ v^2 d\tau^2 - 2 v_i d\tau dx^i \right]+2\rho P d\tau^2\\&\nonumber
-\y \Bigl[\left(4\rho+ 2 P\right) d\Omega^2_2-(1+\rho)  {v_i v_j} dx^i dx^j\\&\nonumber+{\left(\rho^2-1\right)} \left(\nabla^2 v_i+
  R_{j i} v^j \right) d\tau dx^i +2 v_i d\rho dx^i -\left(v^2 + 2 P\right){d\tau d\rho}  + 2    \left(1 + {\rho}\right) \phi_{i}^{(\y)} d\tau dx^i\Bigr]  \\&\nonumber
 +\y^2 \left(4\rho^2d\Omega_2^2 + 2 g_{\rho i}^{(\alpha^2)} dx^id\rho + g_{ij}^{(\y^2)} dx^idx^j\right) + \ldots.
\end{align}
On the spacelike slice at $\rho=-1$, the internal geometry is conformally equal to
\beq
ds_{3d}^2= \left(\frac{1}{\alpha} + 1\right)d\tau^2 + \left(1+2 \alpha\right)^2 d\Omega^2_2~,
\eeq with a conformal factor
\beq
1 - 2 \alpha P + \mathcal{O}(\alpha^2)~.
\eeq
Similarly to the timelike case, for this metric to solve the Einstein equation through $\mathcal{O}(\alpha^0)$, $(v_i,P)$ are required to solve the incompressible Navier-Stokes equation:
\beq
\p_\tau v^i + v_j\nabla_{S^2}^j v^i + \nabla_{S^2}^i P - \nu_{sf} \left(\nabla_{S^2}^2v^i + R^{i}_{j}v^j\right)=0
\eeq
with
\beq
\nu_{sf} = -1~.
\eeq
Note again that the viscosity $\nu_{sf}$ has changed sign from the fluid on the timelike slices. By integrating the ${\cal G}_{\tau\tau}=0$ condition at $\mathcal{O}(\alpha^0)$ over the sphere, we again find the constraint that
\beq
\p_\tau\left(\int v^2 d\Omega_2\right)=-2\p_\tau \left(\int P d\Omega_2\right)~.
\eeq
Hence the structure of the Einstein equation with positive cosmological constant on the timelike slice near the cosmological horizon with the specified boundary conditions is closely mimicked in this spacelike context.

\subsection{Pushing the Spacelike Slice to $\mathcal{I}^+$}

We now wish to push the spacelike slice all the way to $\mathcal{I}^+$ and study the constraints on $\omega$. We impose fast-falling Dirichlet boundary conditions at~$\mathcal{I}^+$ and no incoming flux from the Northern patch.

Reverting to static patch $(r,t)$-coordinates, the solutions $\Phi_v(r,t) = e^{-i\omega t} \varphi_v(r)$ analogous to~(\ref{waves}) and~(\ref{waves2}) near $\mathcal{I}^+$ were computed in~\cite{Anninos:2011jp} and are given by:
\begin{align}
\varphi_v^{-}&=\left(\frac{r^2}{\ell^2}-1\right)^{-i\omega\ell/2}\left(\frac{r}{\ell}\right)^{i\omega}{_2}F_1\left(a_1;b_1;c_1;\frac{\ell^2}{r^2}\right)~,\\
\varphi_v^{+}&=\left(\frac{r^2}{\ell^2}-1\right)^{-i\omega\ell/2}\left(\frac{r}{\ell}\right)^{-1+i\omega\ell}{_2}F_1\left(a_2;b_2;c_2;\frac{\ell^2}{r^2}\right)~,
\end{align}
with
\begin{eqnarray}
a_1 &=& \frac{1}{2}(1+l-i\omega\ell)~, \quad b_1 = \frac{1}{2}(-l-i\omega\ell)~, \quad c_1 = \frac{1}{2}~;\\
a_2 &=& \frac{1}{2}(1-l-i\omega\ell)~, \quad b_2 = \frac{1}{2}(2+l-i\omega\ell)~, \quad c_2 = \frac{3}{2}~.
\end{eqnarray}
As we approach $\mathcal{I}^+$ in the limit $r \to \infty$ we find $\varphi_v^{-} \sim 1$ and $\varphi_v^{+}\sim\ell/r$. Note that $(c_2-a_2-b_2) = i\omega\ell$.
In order to eliminate deformations of the conformal metric at $\mathcal{I}^+$ we switch off the slow falling mode $\varphi_v^{-}$.

Our task is to eliminate the incoming Northern flux upon turning on $\varphi_v^{+}$. It is not hard to see that this will require $(c_2-a_2-b_2) = i\omega\ell$ to be an integer. We must further ensure that metric fluctuations are analytic for $i\omega\ell \in \mathbb{Z}^+$. To achieve this, we must eliminate the logarithmic term in the hypergeometric identity (\ref{logexpand}). This implies that either
\begin{equation}
a_2=-n_1 \quad\text{ or}\quad b_2=-n_2~, \quad n_i = 0,1,2,\dots
\end{equation}
It turns out that of the two possibilities, only the second one is sufficient to eliminate the Northern incoming flux. In the first case, we have to impose a further inequality
$n_1\ge l$, whose origin is explained in appendix \ref{HgeometricIdentity}. Hence, defining $n_1\equiv l+\tilde{n}_1,~\tilde{n}_1=0,1,2,\ldots$ and imposing no further condition on the integer $n_2$, we obtain the following conditions:
\begin{eqnarray}
\label{fqnm1} \omega_{n}^{\mathcal{I}^+} \ell &=& -i(2\tilde{n}_1 + 1+l)~,\quad \tilde{n}_1 = 0,1,2,\dots\\
\label{fqnm2} \omega_{n}^{\mathcal{I}^+} \ell &=& -i(2n_2 + 2+l)~,\quad n_2 = 0,1,2,\dots
\end{eqnarray} which combines into one single tower of modes
\begin{equation}
\omega_{n}^{\mathcal{I}^+} \ell = -i(n+l+1)~,\quad n = 0,1,2,\dots
\end{equation}
Curiously, and perhaps interestingly, this is exactly the same set of modes as the quasinormal mode spectrum~(\ref{qnm}) in the Southern patch.

\subsection{Topological Black Holes in AdS$_4$}

In fact, the above calculation is mathematically equivalent to the computation of quasinormal modes for the massless topological black hole in AdS$_4$ \cite{Birmingham:1998nr,Birmingham:2006zx,Emparan:1999gf} (see also \cite{Horowitz:2009wm}) upon continuing $k_V^2 \to - k_V^2$. This is due to the fact that the metric of the massless topological AdS$_4$:
\begin{equation}\label{topads}
ds^2 = - \left( -1 + \left(\frac{\tilde{r}}{\ell_{AdS}} \right)^2 \right) d\tilde{t}^2 + \left( -1 + \left(\frac{\tilde{r}}{\ell_{AdS}} \right)^2 \right)^{-1} d\tilde{r}^2 + \tilde{r}^2 d \mathcal{H}_2^2~
\end{equation}
is related to the static patch metric by an analytic continuation. The two-dimensional space: $d\mathcal{H}_2^2 = ( d\xi^2 + \sinh^2 \xi d\tilde{\phi}^2)$ is the standard metric on the hyperbolic two-manifold. The analytic continuation from the static patch metric (\ref{static}) to the above metric is given by:
\begin{equation}
\ell \to i \ell_{AdS}~, \quad t \to i \tilde{t}~, \quad r \to i \tilde{r}~, \quad \theta \to i \xi~,\quad \phi \to \tilde{\phi}~.
\end{equation}
An observer in the massless topological AdS$_4$ geometry observes a Hawking temperature given by $T = 1/2\pi\ell_{AdS}$. If one considers compact quotient of $\mathcal{H}_2$, there is a finite entropy proportional to $(\ell_{AdS}/\ell_{P})^2$ associated with the horizon at $\tilde{r}= \ell_{AdS}$. We also expect such mathematical similarities to hold between the boundary correlators near the boundary of topological AdS$_4$ black holes and those near $\mathcal{I}^+$ (using the boundary conditions discussed above). It is also interesting to note that at the non-linear level one can add negative energy to (\ref{topads}) and create spherically symmetric asymptotically AdS$_4$ topological black holes. This occurs up to a critical negative mass, for which one finds a solution interpolating between AdS$_4$ near the boundary and AdS$_2\times \mathcal{H}^2$ \cite{Birmingham:1998nr}. Similarly, adding sufficient mass to the static patch leads to the Nariai solution which interpolates between dS$_4$ near $\mathcal{I}^+$ and dS$_2 \times S^2$. We hope to further uncover this map in future work, as it may provide insight into the nature of the static patch observer.

\section*{Acknowledgments}

It is a great pleasure to thank Tatsuo Azeyanagi, Frederik Denef, Willy Fischler, Tom Hartman, Sean Hartnoll, Diego Hofman, Vyacheslav Lysov, Loganayagam Ramalingam  and Andy Strominger for useful discussions. D.A. is particularly grateful to Frederik Denef for countless hours of insightful discussions. T.A. would like to thank the Stanford Institute for Theoretical Physics for their hospitality during the completion of this work. The research has been partly funded by the NSF Grant NSF PHY05-51164, the Center for the Fundamental Laws of Nature, DOE grant DE-FG02-91ER40654 and the John Templeton Foundation.

\appendix

\section{Scalar Perturbations}\label{scalar}

Gravitational perturbations in static dS$_4$ consist of a scalar and vector mode of the $SO(3)$ of the two-sphere. There are no tensor modes in four dimensions.
Scalar harmonic perturbations can be reduced to a single Ishibashi-Kodama master field \cite{Kodama:2000fa,Kodama:2003jz}, which obeys the same effective equation as that of the vector perturbations except that the angular number $l$ begins at $l=0$ (instead of $l=1$ in the vector case).

An incompressible fluid requires a divergenceless velocity field $v_i$. The scalar harmonic allows only the possibility $v_i \propto {\mathcal{ S }}_i \equiv -\nabla_i {\mathcal{ S }}$. $\mathcal{ S }$ is the scalar harmonic on the sphere which satisfies:
\begin{equation}
\left(\nabla_{S^2}^2+k_S^2 \right) {\mathcal{ S } } = 0~,  \quad k_S^2 = l(l+1)~, \quad l=0,1,2,\ldots
\end{equation}
Imposing incompressibility leads to
\begin{equation}
\nabla_{S^2}^i v_i \propto \nabla_{S^2}^2 {\mathcal{ S  }} = - k_S^2 {\mathcal{ S }}
\end{equation}
which vanishes for $k_S=0$, i.e. the spherically symmetric mode $l=0$. In this case $v_i = 0$ and we are left with a trivial fluid.

Thus, in the case of an incompressible fluid, the scalar mode only consists of trivial fluids. It would be interesting to investigate the case of a compressible fluid which allows for sound modes.\footnote{In \cite{Bredberg:2010ky}, it was shown for planar horizons that the speed of sound for the scalar sound mode goes to infinity and effectively decouples from the non-relativistic fluid sector.}

\section{The $l = 1$ Vector Perturbation}\label{smallrot}

Gravitational vector perturbations with $l = 1$ differ from $l > 2$. We follow the discussion in  \cite{Kodama:2000fa,Kodama:2003jz}.
In addition to equation~\ref{vecharmoniceq}, spherical vector harmonics satisfy the following equation:
\begin{equation}
\left(\nabla^2_{S^2} + k_V^2 - 3 \right) {\mathcal V}_{ij}=0\quad,\quad {\mathcal V}_{ij} \equiv -\frac{1}{2k_V} \left[D_i {\mathcal V}_j +D_j {\mathcal V}_i\right]~,
\end{equation} where $D_i\equiv \left(\nabla_{S^2}\right)_i$.
For $k_V^2-3 \leq 0$ it can be shown that ${\mathcal V}_{ij}$ vanishes and therefore, ${\mathcal V}^i$ must be a Killing vector on the sphere. In this case, we parametrize the perturbations as:
\begin{equation}
\delta g_{ai} = r f_a {\mathcal V}_i,
\end{equation} where $x^a=\left\{t,r\right\}$ and $x^i=\left\{\theta,\phi\right\}$.
Given that ${\mathcal V}_{ij}=0$ implies $\delta g_{ij}=0$, we can no longer fix the gauge freedom by imposing $\delta g_{ij}=0$.
Instead, we will fix the gauge $f_r=0$.  From \cite{Kodama:2003jz},
using the only gauge invariant object $F_{ab}$:
\begin{equation}
r^{-1}F_{ab} =D_a \left(\frac{f_b}{r}\right)- D_b \left( \frac{f_a}{r}\right)~,~~ D_a\equiv \left(\nabla_{g^{(2)}}\right)_a
\end{equation}
the Einstein equation imply that $F_{ab}$ is:
\begin{equation}
F_{ab} = \frac{3 J}{r^3} \epsilon_{ab}
\end{equation}
with $J$ constant. In the case of a perturbation about a spherically symmetric black hole, this corresponds to giving it a small amount of angular momentum~\cite{Kodama:2003jz}.

We can study the $r$ dependence of the metric components. Given that the connection term will drop out due to anti-symmetrization, we can replace the covariant derivatives with ordinary derivatives and find:
\begin{equation}
-\frac{3J}{r^4} = \partial_r \left(\frac{f_t}{r}\right)~.
\end{equation}
The above integrates to
\begin{equation}
f_t(r) =J\left(\frac{1}{r^2}- \frac{r}{r_c^3} \right)~,
\end{equation}
where we have set Dirichlet boundary conditions on $\delta g_{it}$ at $r=r_c$. Hence, the perturbed metric reads
\begin{equation}
ds^2=ds_0^2+
J\left[\frac{1}{r} -\frac{r^2}{r_c^3}
\right]{\mathcal V}_i(\theta,\phi) dt dx^i
\end{equation}
with ${\mathcal V}_i$ a Killing vector on $S^2$.


\section{Mind the Gap }\label{mindgap}
Looking at the last two plots in figure~\ref{flowpoles}, we see that as we move the cutoff surface away from the cosmological horizon, the zeroes for $l=9$ and $l=10$ modes jump discontinuously. We would like to describe how this behavior arises. Recall that our goal was to find the lowest lying zero of $\delta g_{i\tau}^{out}(\rho=1)$ as a function of $\omega$ on the negative imaginary axis as we move $r_c$.
\begin{figure}[ht!]
\centering\
\includegraphics[width=0.48\textwidth]{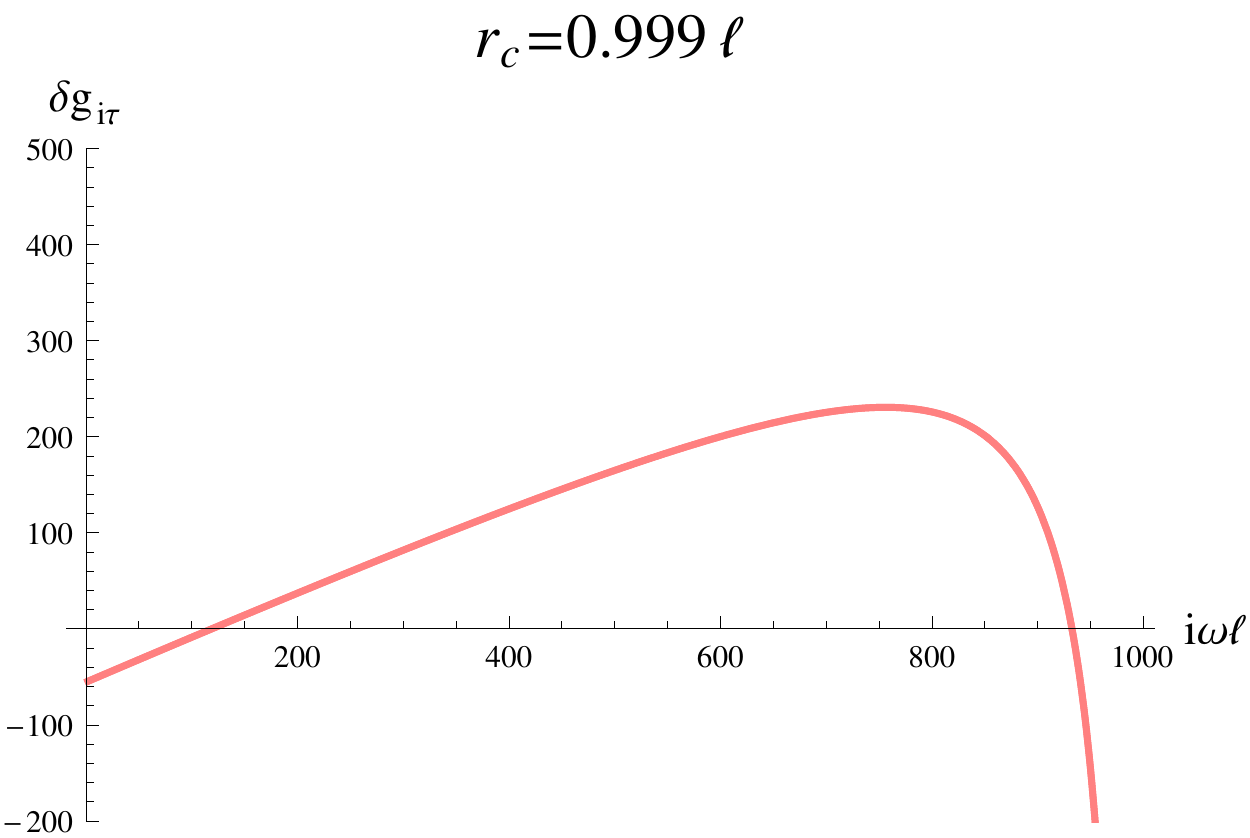}
\includegraphics[width=0.48\textwidth]{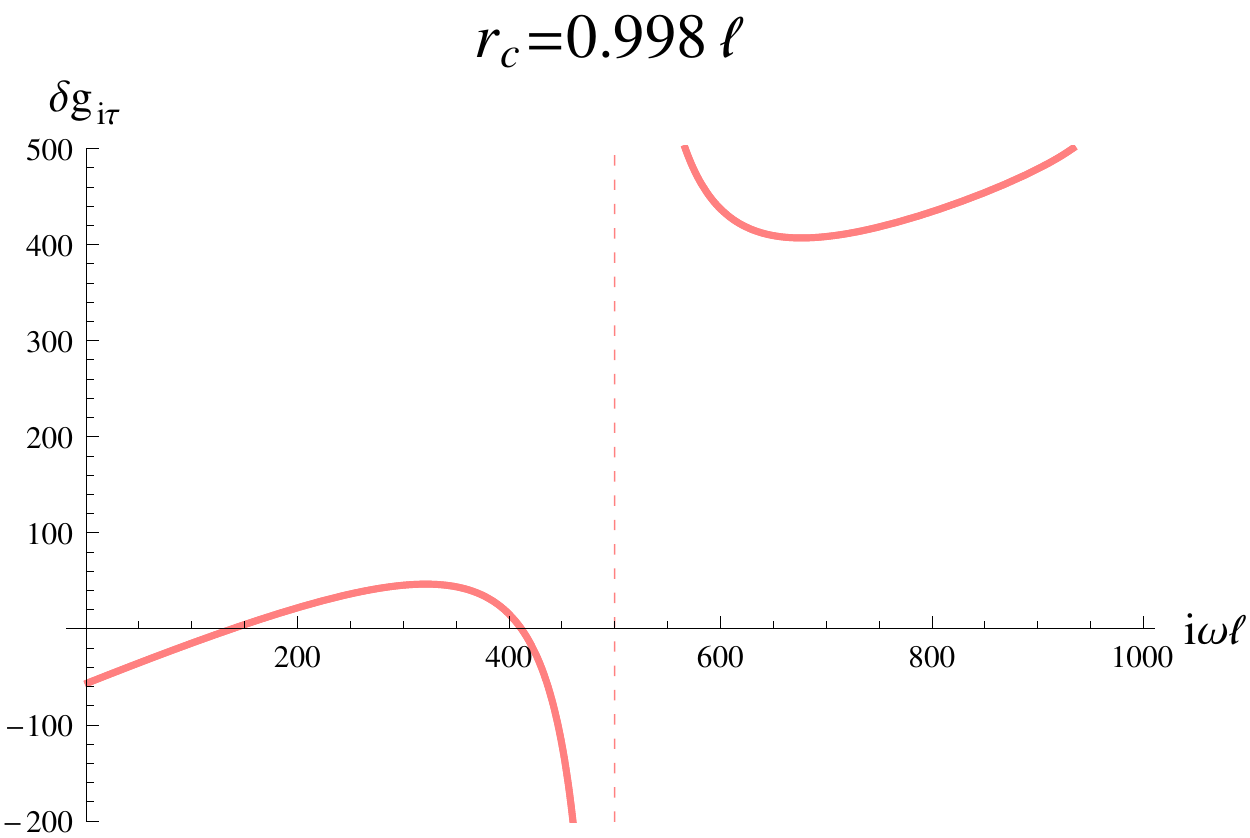}
\includegraphics[width=0.48\textwidth]{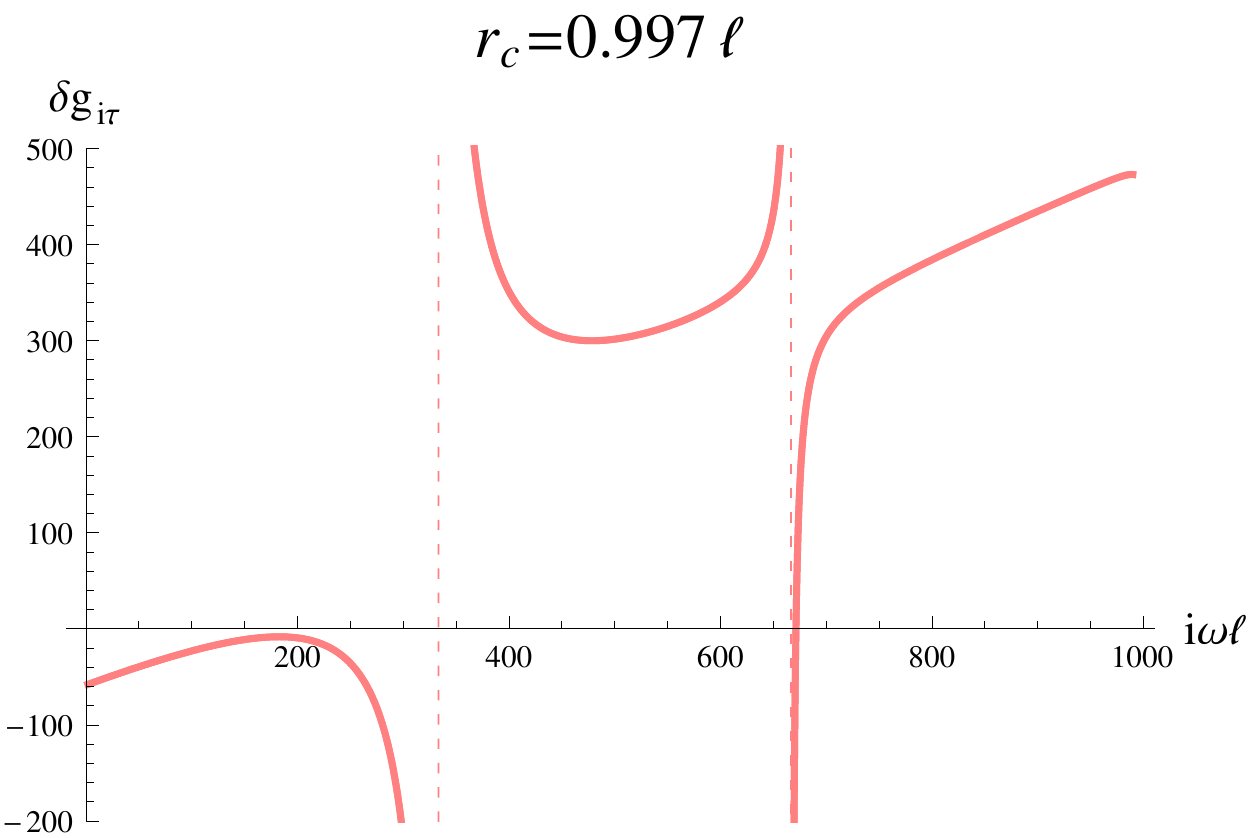}
\includegraphics[width=0.48\textwidth]{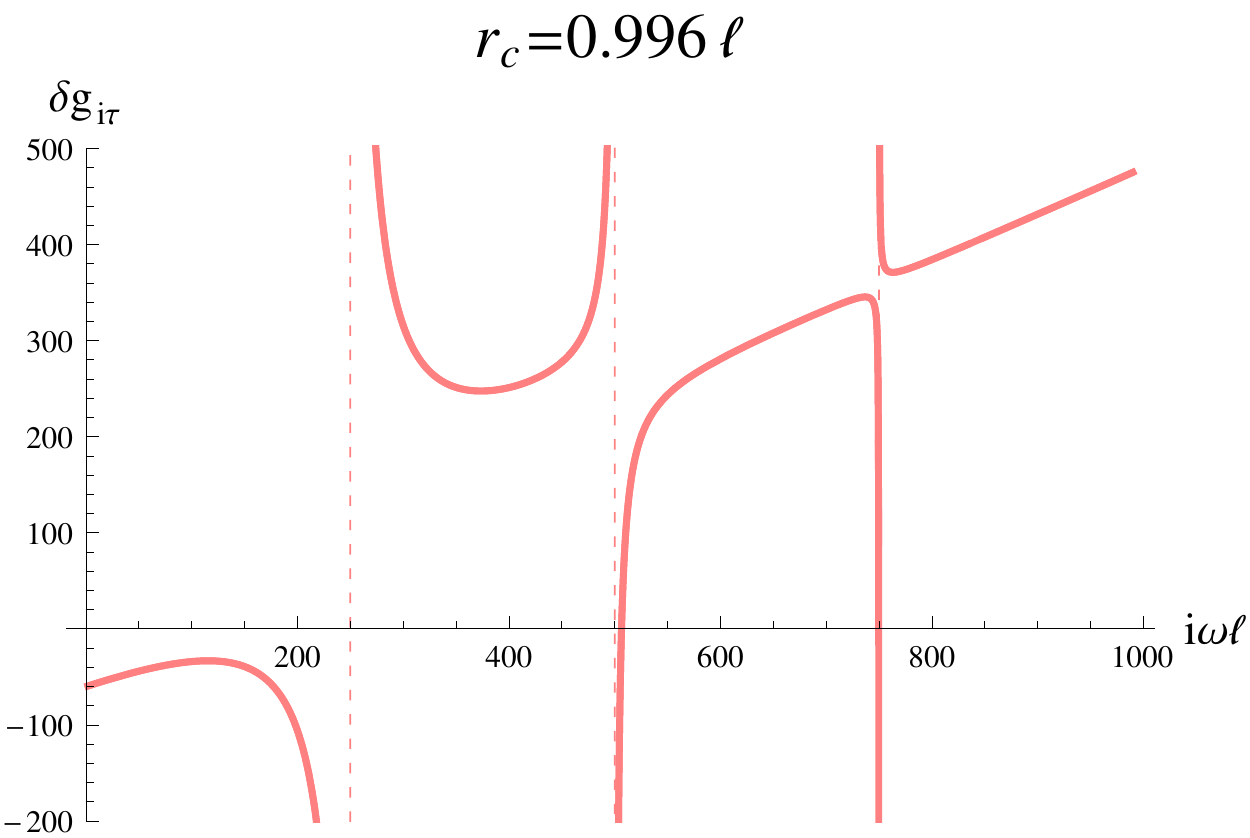}
\caption{Flow of $\delta g_{i\tau}^{out}(\rho=1)$ for $l=10$ as we move $r_c$ away from the cosmological horizon. We wish to find frequencies that make this function vanishes. As we vary $r_c$ the zero jumps. }\label{mindfig}
\end{figure}
Notice in figure~\ref{mindfig} that as we move $r_c$ to smaller values, the lowest lying zero disappears and the new lowest lying zero is at a finite distance away in frequency space.

\section{Hypergeometric Gymnastics}\label{HgeometricIdentity}

For $(c-a-b)$ positive integer the following relation holds \cite{bateman}:
\begin{multline}\label{logexpand}
{_2}F_1\left(a;b;c;z\right)=\frac{\Gamma(c-a-b)\Gamma(c)}{\Gamma(c-b)\Gamma(c-a)}\sum_{n=0}^{c-a-b-1}\frac{(a)_n(b)_n}{(1+a+b-c)_n n!}(1-z)^n\\
+(z-1)^{c-a-b}\frac{\Gamma(c)}{\Gamma(a)\Gamma(b)}\sum_{n=0}^{\infty}\frac{(c-b)_n(c-a)_n}{n!(n+c-a-b)!}\left[k_n-\log(1-z)\right](1-z)^n~,
\end{multline}
where $(a)_n\equiv\Gamma(a+n)/\Gamma(a)$ are the Pochhammer symbols and
\begin{equation}
k_n=\psi(n+1)+\psi(n+1+c-a-b)-\psi(n+c-a)-\psi(n+c-b)
\end{equation}
with $\psi(z)=d\log\Gamma(z)/dz$.

We treat the case covered by equation~(\ref{logexpand}) as it is relevant to the text. To eliminate the log terms, we require $1/\Gamma(a)\Gamma(b)=0$. In the case of $a=-n_1,~n_1=0,1,2,\ldots$, if $c-b>0$ (which in the spacelike case becomes $n_1 \ge l$), then the whole second sum vanishes and we get \cite{WangGuo}:
\begin{equation}
{_2}F_1\left(a;b;c;z\right) =
\frac{\Gamma(c-a-b)\Gamma(c)}{\Gamma(c-b)\Gamma(c-a)}
{_2}F_1\left(a;b;1-c;1-z\right)
\end{equation} which goes to a constant as $r\rightarrow 1$ (here $z=\ell^2/r^2$) and translates into the modes with no incoming flux from the Northern patch. So, these are the modes we want.\footnote{In the case of $b=-n_2,~n_2=0,1,2,\ldots$, the analogous inequality in the first regime is $c-a>0$.
However, our parameters imply that $c-a>0$ is always true. Thus we are always in this first regime which gives the modes that we want and there is no further restriction on $n_2$.}

In the other regime $c-b\le0$, the first sum vanishes completely due to the gamma function's poles. Naively, we would think that the whole expression is zero, however, the second term contains a divergent term $\psi(n+c-b)$ in the sum that will cancel out the divergence in $1/\Gamma(a)$ for $n \le -(c-b)$. then we actually have
\begin{equation}
\frac{\psi(n+c-b)}{\Gamma(a)} \sim
\Gamma(-a+1)
\end{equation} leading to\cite{WangGuo}:
\begin{equation}
{_2}F_1\left(a;b;c;z\right)
\sim
\frac{\Gamma(c)\Gamma(-a+1)}{\left(c-a-b\right)!\Gamma(b)}
\left(1-z\right)^{c-a-b}{_2}F_1\left(c-b;c-a;1+c;1-z\right).
\end{equation} This implies that the $\varphi_v^+$  tends to $\left(r^2-1\right)^{i\omega/2}$ (as $r\rightarrow 1$) which is an incoming wave from the Northern patch and should be excluded.

\end{document}